\documentclass[showpacs,twocolumn,pra,longbibliography]{revtex4-1}

\usepackage{epsf}
\usepackage{amsmath}
\usepackage{amsfonts}
\usepackage{amssymb}
\usepackage{bm}
\usepackage{bbm}
\usepackage{nicefrac}
\usepackage{color}
\usepackage{maybemath}
\usepackage{pifont}

\usepackage{graphicx}
\usepackage{epsfig}
\usepackage{latexsym}
\usepackage{pifont}

\def\calH{{\mathcal{H}}}
\def\calL{{\mathcal{L}}}
\def\calP{{\mathcal{P}}}
\def\calQ{{\mathcal{Q}}}

\def\calT{{\mathcal{T}}}
\def\calV{{\mathcal{V}}}
\def\calW{{\mathcal{W}}}

\def\rmT{{\mathrm{T}}}

\def\dd{{\mathrm{d}}}
\def\ii{{\mathrm{i}}}
\def\ee{{\mathrm{e}}}

\newcommand{\dir}{{(\rm{dir})}}
\newcommand{\mix}{{(\rm{mix})}}

\def\vdw{van der Waals}
\def\cp{Casimir--Polder}

\begin{document}

\title{Long--Range Tails in \vdw{} Interactions \\
of Excited--State and Ground--State Atoms}

\newcommand{\addrROLLA}{Department of Physics,
Missouri University of Science and Technology,
Rolla, Missouri 65409-0640, USA}

\author{U. D. Jentschura}
\affiliation{\addrROLLA}

\author{V. Debierre}
\affiliation{\addrROLLA}

\begin{abstract}
A quantum electrodynamic calculation of the 
interaction of an excited-state atom with a ground-state
atom is performed. For an excited reference state and a 
lower-lying virtual state, the contribution to the 
interaction energy naturally splits into a pole term,
and a Wick-rotated term. The pole term is shown to 
dominate in the long-range limit, altering the 
functional form of the interaction from the retarded 
$1/R^7$ \cp{} form to a long-range tail--provided 
by the Wick-rotated term--proportional to 
$\cos[2 (E_m-E_n) \, R/(\hbar c)]/R^2$, where $E_m < E_n$ is the 
energy of a virtual state, lower than the reference state energy $E_n$,
and $R$ is the interatomic separation.
General expressions are obtained which can be applied 
to atomic reference states of arbitrary angular symmetry.
Careful treatment of the pole terms in the 
Feynman prescription for the atomic polarizability 
is found to be crucial in obtaining correct results.
\end{abstract}

\maketitle

%
%
\section{Introduction}
\label{sec1}

Recently, the long-range tails of the
interaction between an excited-state 
and a ground-state 
atom~\cite{DoGuLa2015,SaKa2015,Be2015,MiRa2015,DoGuLa2015,Do2016}
as well as those of the interaction between an excited $2S$ state with a 
conducting wall~\cite{Je2015rapid},
have received considerable attention.
The question behind the investigation concerns 
the existence of long-range tails for excited 
reference states, for which partially 
conflicting results have been obtained in the 
past~\cite{GoMLPo1966,PoTh1995,SaBuWeDu2006}.

In this article, we reconsider the derivation of the 
long-range interaction, with a particular emphasis
on the interaction of an excited-state atom with another
ground-state atom, their separation being large 
compared to the Bohr radius. 
We follow a method that deduces the long-range interaction 
from the scattering amplitude
[see Chap.~85 of Ref.~\cite{BeLiPi1982vol4}].
This method demands the use of the Feynman 
prescription for the Green functions of the 
photon field, and the time-ordered product of 
atomic dipole operators.

We also aim to generalize the recent
treatments in Refs.~\cite{SaKa2015,DoGuLa2015}
to reference states of arbitrary symmetry,
and to clarify the role of the 
virtual-state energy in the calculation of the 
final expressions, without any approximations.
In our formalism, we aim to calculate the 
long-range tails of the \vdw{} and \cp{}
energy shifts on the basis of a unified 
formalism, which can be applied to both ground-state
and excited-state interactions, with atomic 
state of arbitrary symmetry.
The general idea is to use the matching 
of the forward scattering amplitude
from quantum electrodynamics (QED), against the 
effective potential that describes the long-range
interaction.

The paper is organized as follows.
In Sec.~\ref{sec2}, we reconsider the derivation of the 
\vdw{} and \cp{} interaction from first principles,
using the matching of the $S$ matrix 
element with the effective interaction potential.
Applications are discussed in Sec.~\ref{sec3}.
First, in order to check our results and connect them 
to the literature, we rederive the familiar form of the 
ground-state interaction in Sec.~\ref{sec31},
and verify the \vdw{} close-range limit in
Sec.~\ref{sec32}. 
General excited states are treated in Sec.~\ref{sec33},
and the expressions are specialized to 
excited $S$ states in Sec.~\ref{sec34}.
Finally, conclusions are reserved for Sec.~\ref{sec4}.

%
%
\section{Derivation}
\label{sec2}

%
%
\subsection{$\maybebm{S}$--Matrix and Matching with Effective Interaction}
\label{sec21}

We consider two atom in states $\psi_A(\vec r_A)$ and 
$\psi_B(\vec r_B)$ which scatter into states
$\psi'_A(\vec r_A)$ and $\psi'_B(\vec r_B)$ 
under the action of a potential $U(\vec r_A, \vec r_B, \vec R)$.
Here, the absolute electron coordinates are 
$\vec x_A$ and $\vec x_B$; the relative coordinates are 
$\vec r_A = \vec x_A - \vec R_A$ and
$\vec r_B = \vec x_B - \vec R_B$,
where $\vec R_A$ and $\vec R_B$ are the
coordinates of the nucleus. 
Their distance is $\vec R \equiv \vec R_A - \vec R_B$.
We denote the initial state by $i$ 
(atoms are in states $\psi_A$ and $\psi_B$,
respectively) and the final 
state by the subscript $f$
(atoms are in states $\psi'_A$ and $\psi'_B$).
The corresponding $S$-matrix element 
reads as follows~\cite{ItZu1980},
\begin{align}
\label{SApBpAB}
S_{A'B'AB} =& \; -\frac \ii \hbar \,
\int \dd^3 r_A \int \dd^3 r_B \;
\psi_A'^*(\vec r_A) \, \psi_B'^*(\vec r_B) \,
\nonumber\\
& \; \times U(\vec r_A, \vec r_B, \vec R) \,
\psi_A(\vec r_A) \, \psi_B(\vec r_B) \\
&\times
\int \dd t \, \exp\left[ -\frac \ii \hbar \, (E_1 + E_2 - E'_1 - E'_2) \, t\right] 
\nonumber\\
=& \; -\frac \ii \hbar \, T \, 
\int \dd^3 r_A \int \dd^3 r_B \;
\psi_A'^*(\vec r_A) \, \psi_B'^*(\vec r_B) \,
\nonumber\\
& \; \times U(\vec r_A, \vec r_B, \vec R) \,
\psi_A(\vec r_A) \, \psi_B(\vec r_B) \,,
\end{align}
where we have assumed energy conservation
($E_1 + E_2 = E'_1 + E'_2$) and 
denoted the (long) time interval over which the 
transition from initial to final state occurs, 
as $\int \dd t = T$.
The matching of the effective
perturbative  Hamiltonian $H_{\rm eff}$ and the 
$S$ matrix element thus is 
\begin{align}
\label{match}
\langle \psi'_A, \psi'_B | H_{\rm eff} 
| \psi_A, \psi_B \rangle =& \;
\langle \psi'_A, \psi'_B | U( \vec r_A, \vec r_B, \vec R ) 
| \psi_A, \psi_B \rangle 
\nonumber\\[0.1133ex]
=& \; \frac{\ii \hbar}{T} \, S_{A'B'AB} \,.
\end{align}
On the level of a scattering matrix element,
the matching is obtained 
in an ``averaged'' sense,
where the ``averaging'' (i.e., the integration)
occurs over the wave functions 
of the initial and final states of the two-atom system.
In the following, we shall concentrate on 
forward scattering, i.e.,
$| \psi_{A'} \rangle = | \psi_A \rangle $,
$| \psi_{B'} \rangle = | \psi_B \rangle $.

%
%
\subsection{Interaction Hamiltonian}
\label{sec22}

We are inspired by the derivation outlined in 
Chap.~85 of Ref.~\cite{BeLiPi1982vol4}.
We shall use time-dependent QED perturbation theory,
where the interaction is formulated in the 
interaction picture~\cite{ItZu1980,MoPlSo1998}.
This means that the second-quantized operators in the interaction
Hamiltonian have a time dependence 
which is generated by the action of the 
free Hamiltonian~\cite{JeKe2004aop}.
We shall use a second-quantized approach for the 
operators describing the electromagnetic field,
so that a time-ordered product of the 
four-vector potential operators results in the 
Feynman propagator of the photon~\cite{ItZu1980}.
For the position operators of the atomic 
electrons, though, we use a first-quantized approach, 
i.e., we treat these on the level of 
quantum mechanics, without the 
introduction of fermion creation and annihilation operators.

The interaction Hamiltonian in the dipole approximation then is
\begin{align} \label{eq:IntPot}
V(t) =& \; -\vec E(\vec x_A, t) \cdot \vec d_A(t)
-\vec E(\vec x_B, t) \cdot \vec d_B(t) 
\nonumber\\
\approx & \; -\vec E(\vec R_A, t) \cdot \vec d_A(t)
-\vec E(\vec R_B, t) \cdot \vec d_B(t) \,,
\end{align}
where $\vec d_i = e \, \vec r_i$ is the 
dipole operator for atom $i$ (for atoms with more 
than one electron, one has to sum over all the electrons
in the atoms $i=A,B$). The $\vec R_A$ and $\vec R_B$
are the positions of the atomic nuclei.
A clarifying remark is in order: In the standard formulation of quantum 
electrodynamics, one would use the 
interaction Hamiltonian density $\calH = j^\mu \, A_\mu$,
where $j^\mu = \hat{\overline \psi} \, \gamma^\mu \, \hat{\psi}$
is the fermionic current operator, 
$\gamma^\mu$ are the Dirac $\gamma$ matrices, 
and the $A_\mu$ is the four-vector potential~\cite{ItZu1980,MoPlSo1998}.
The fermionic field operator $\hat\psi$ contains the 
fermionic creation and annihilation operators.
However, in the nonrelativistic limit,
one may renounce on the quantization of the 
fermion field, and treat the electronic degrees
of freedom using first quantization~\cite{CrTh1984,JeKe2004aop}.

\begin{figure} [t]
\begin{center}
\begin{minipage}{0.99\linewidth}
\begin{center}
\includegraphics[width=0.8\linewidth]{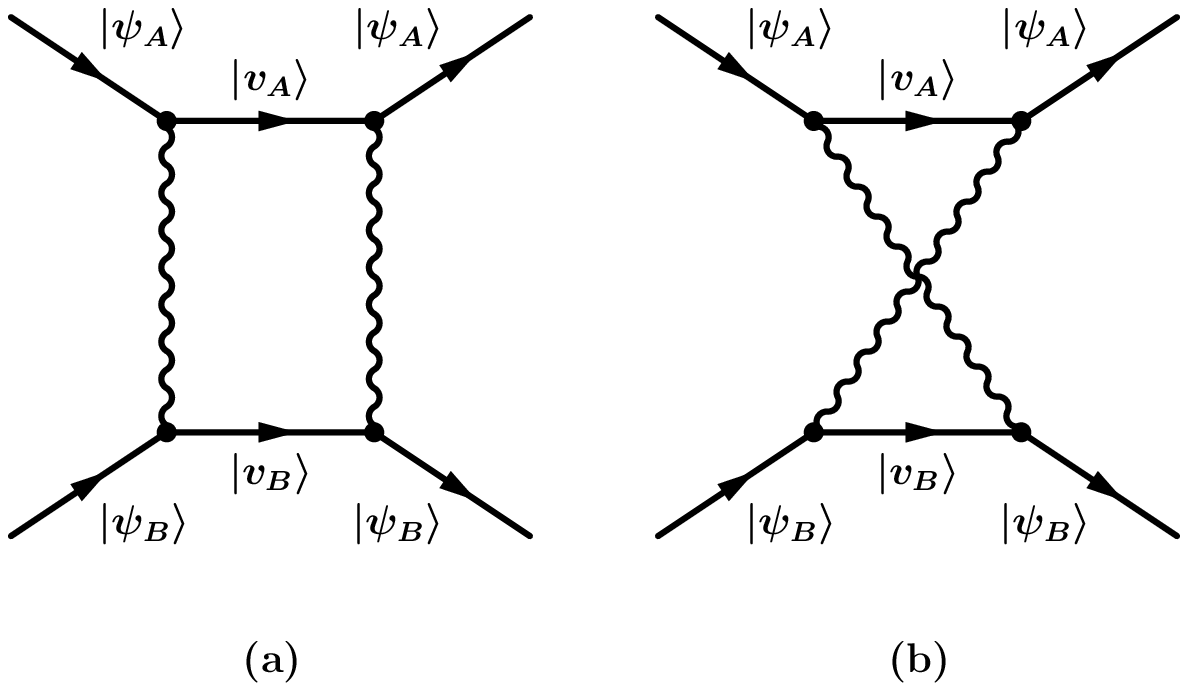}
\caption{\label{fig1}
Feynman diagrams for the excited-state long-range 
interaction of an atom in state $|\psi_A\rangle$
(excited) with a ground-state atom $B$, in state $|\psi_B\rangle$.
Figure (a) is the ladder diagram, while~(b) displays the 
crossed-ladder graph.
The power of using the Feynman propagator in the 
calculation lies in the fact that all the different time orderings
of the electron-photon vertices,
which are otherwise relevant to time-ordered perturbation theory~\cite{CrTh1984},
can be summarized in only two diagrams.}
\end{center}
\end{minipage}
\end{center}
\end{figure}

The fourth-order contribution to the 
$S$-matrix is (the full matrix, not a single element)
\begin{align}
S^{(4)} =& \; \frac{(-\ii)^4}{4!\,\hbar^4}
\int \dd t_1 \int \dd t_2 \int \dd t_3 \int \dd t_4
\nonumber\\
& \; \times {\bf T}[ V(t_1) V(t_2) V(t_3) V(t_4) ] \,,
\end{align}
where ${\bf T}$ denotes the time ordering of 
all operators, pertaining both to the atomic as well 
as the field degrees of freedom.
According to the Wick theorem,
the time-ordered product is equal to the 
normal ordered product, plus all contractions.
We need to calculate the fourth-order $S$ matrix element 
$\langle \psi,0 |  S^{(4)} | \psi,0 \rangle$
for forward scattering of the atomic reference 
state $|\psi\rangle = |\psi_A, \psi_B\rangle
= |\psi_A\rangle \otimes | \psi_B\rangle$
with the vacuum $| 0 \rangle$ of the electromagnetic field
(the product state is $|\psi, 0\rangle$).
After the subtraction of terms which pertain to the 
self-energies of the atoms, one obtains four 
contributions which are proportional to 
(${\rm T}$ denotes the time ordering of 
 dipole operators)
\begin{widetext}
\begin{subequations}
\begin{align}
C_1 \equiv & \; 
\langle \psi_A | \rmT \; d_{Ai}(t_1) \, d_{Ak}(t_3) | \psi_A \rangle \,
\langle \psi_B | \rmT \; d_{Bj}(t_2) \, d_{B\ell}(t_4) ) | \psi_B \rangle \,,
\\
C_2 \equiv & \; 
\langle \psi_A | \rmT \; d_{Ai}(t_1) \, d_{A\ell}(t_4) | \psi_A \rangle \,
\langle \psi_B | \rmT \; d_{Bj}(t_2) \, d_{Bk}(t_3) | \psi_B \rangle \,,
\\
C_3 \equiv & \; 
\langle \psi_B | \rmT \; d_{Bi}(t_1) \, d_{B\ell}(t_4) | \psi_B \rangle \,
\langle \psi_A | \rmT \; d_{Aj}(t_2) \, d_{Ak}(t_3) | \psi_A \rangle \,,
\\
C_4 \equiv & \; 
\langle \psi_B | \rmT \; d_{Bi}(t_1) \, d_{Bk}(t_3) | \psi_B \rangle \,
\langle \psi_A | \rmT \; d_{Aj}(t_2) \, d_{A\ell}(t_4) | \psi_A \rangle \,.
\end{align}
\end{subequations}
\end{widetext}
Contributions $C_2$ and $C_4$ correspond to the 
crossed-ladder diagram (in the language of Feynman diagrams,
see Fig.~\ref{fig1}), whereas 
$C_1$ and $C_3$ correspond to the two-photon ladder exchange.
The contributions of atoms $A$ and $B$ to the 
atomic reference state are denoted as $| \psi_A \rangle$ and 
$| \psi_B \rangle$, respectively.
All terms $C_1$, $C_2$, $C_3$, and $C_4$
lead to equivalent contributions, and we finally 
arrive at ($\calT$ denotes the time 
ordering of field operators)
\begin{multline}
\label{S4_expr1}
\langle \psi, 0 | S^{(4)} | \psi, 0 \rangle 
= \frac{1}{2\,\hbar^4} 
\int \dd t_1 \int \dd t_2 \int \dd t_3 \int \dd t_4 \,
\\[0.1133ex]
\times \langle 0 |
\calT\left[ E_i(\vec R_A, t_1) \, E_j(\vec R_B, t_2) \right]
| 0 \rangle 
\\[0.1133ex]
\times \langle 0 |
\calT\left[ E_k(\vec R_A, t_3) \, E_\ell(\vec R_B, t_4) \right]
| 0 \rangle 
\\[0.1133ex]
\times \left< \psi_A \left|
\rmT \; d_{Ai}( t_1) \, d_{Ak}( t_3) 
\right| \psi_A  \right> 
\\[0.1133ex]
\times \left< \psi_B \left|
\rmT \; d_{Bj}( t_2) \, d_{B\ell}( t_4) 
\right| \psi_B \right> \,.
\end{multline}

%
%
\subsection{Temporal Gauge and Propagator}

The time-ordered product of electric-field operators
can be evaluated as follows,
\begin{equation}
D^E_{ik}(x_1 - x_2) = 
\left< 0 \left|
\calT\left[ E_i(\vec R_A, t_1) \, E_k(\vec R_B, t_2) \right]
\right| 0 \right> \,.
\end{equation}
With $\vec E = -\partial_t \vec A$, we have
for the ``electric-field propagator'' $D^E_{ik}(x_1 - x_2)$,
\begin{multline}
D^E_{ik}(x_1 - x_2) 
= \left( -\frac{\partial}{\partial t_1} \right) \,
\left( -\frac{\partial}{\partial t_2} \right) \,
\\
\times \left< 0 \left|
\calT\left[ A_i(\vec R_A, t_1) \, A_k(\vec R_B, t_2) \right]
\right| 0 \right> \,.
\end{multline}
One can relate the time-ordered product of field
operators to the photon propagator,
\begin{equation}
\left< 0 \left| \calT A_\mu(x) \, A_\nu(x') \right| 0 \right> = 
-\ii \, D_{\mu\nu}(x - x') \,.
\end{equation}
We resort to the Fourier representation for the temporal gauge 
(also known as the Weyl gauge,
with vanishing scalar component $D_{00} = 0$ 
and $D_{i0} = D_{0i} = 0$).
According to Eq.~(76.14) of Ref.~\cite{BeLiPi1982vol4},
one has
\begin{equation}
D_{ik}(\omega,\vec{k}) = 
-\frac{\hbar}{4 \pi \epsilon_0c^2} \,
\frac{1}{\left(\frac{\omega}{c}\right)^2 - \vec k^2 + \ii \epsilon} \, 
\left( \delta_{ik} - c^2\frac{k_i \, k_k}{\omega^2} \right) \,.
\end{equation}
According to Eq.~(76.16) of Ref.~\cite{BeLiPi1982vol4}, 
the propagator in the mixed frequency-position representation is given by
\begin{equation}
D_{ik}(\omega, \vec R) =
- \left( \delta_{ik} + c^2\frac{\nabla_i \, \nabla_k}{\omega^2} \right) \,
D(\omega, \vec R) \,,
\end{equation}
where
\begin{equation}
D(\omega, \vec R) =
- \frac{\hbar}{4 \pi \epsilon_0 c^2} \,
\frac{\ee^{\ii \sqrt{\omega^2 + \ii \epsilon} \, R/c}}{R} \,,
\end{equation}
and $\epsilon$ is an infinitesimal parameter
used in the frequency-coordinate representation of the
the Feynman propagator.
In the following, we shall use the nonstandard definition
\begin{equation}
\label{nonstandard}
|\omega| \equiv \sqrt{\omega^2 + \ii \epsilon}
\end{equation}
for complex photon frequency $\omega$.
We carry out the differentiations with the result,
\begin{multline}
\nabla_i \, \nabla_k \frac{\ee^{\ii \frac{|\omega|}{c} R}}{R} =
\left(\frac{\omega}{c}\right)^2 \, \delta_{ik} \,
\left( - \frac{c^2}{\omega^2 \, R^2}
+ \frac{\ii c}{|\omega| \, R} \right) \,
\frac{\ee^{\ii \frac{|\omega|}{c} R}}{R} 
\\[0.1133ex]
+ \left(\frac{\omega}{c}\right)^2 \, \frac{R_i \, R_k}{R^2} \,
\left( \frac{3c^2}{\omega^2 \, r^2}
- \frac{3 \ii c}{|\omega| \, R}
- 1 \right) \,
\frac{\ee^{\ii \frac{|\omega|}{c}R}}{R} \,.
\end{multline}
The temporal gauge photon propagator 
in the mixed representation becomes
\begin{subequations} 
\label{sum2}
\begin{multline} 
D_{ik}(\omega, \vec R) =
\frac{\hbar}{4 \pi \epsilon_0c^2}\left[
\delta_{ik} 
\left( 1 + \frac{\ii c}{|\omega| R} 
- \frac{c^2}{\omega^2 R^2} \right) \,
\frac{\ee^{\ii \frac{|\omega|}{c} R}}{R}
\right.
\\
\left.
+ \frac{R_i \, R_k}{R^2} \,
\left(- 1 - \frac{3 \ii c}{|\omega| \, R}
+ \frac{3c^2}{\omega^2 \, R^2} \right) \,
\frac{\ee^{\ii \frac{|\omega|}{c} R}}{R} \right] \,
\\[0.1133ex]
= \frac{\hbar}{4 \pi \epsilon_0 c^2}\;
\left[ \alpha_{ik} + \beta_{ik} \,
\left( \frac{\ii c}{|\omega| R} -
\frac{c^2}{\omega^2 \, R^2} \right) \right] \,
\frac{\ee^{\ii \frac{|\omega|}{c} R}}{R} \,,
\end{multline}
where 
\begin{multline}
\alpha_{ik} = \delta_{ik} - \frac{R_i \, R_k}{R^2} \,,
\qquad
\beta_{ik} = \delta_{ik} - 3 \frac{R_i \, R_k}{R^2} \,.
\end{multline}
\end{subequations} 
The photon propagator, which is the propagator for the 
vector potential $\vec A$, can be translated into the 
propagator for the electric field by differentiation
with respect to time,
\begin{align}
D^E_{ik}(x_1 - x_2) =& \;
\frac{\partial}{\partial t_1} 
\frac{\partial}{\partial t_2} 
\langle 0 |
\calT A_i(\vec R_A, t_1) \, A_k(\vec R_B, t_2) 
| 0 \rangle
\nonumber\\[0.1133ex]
=& \;
\frac{\partial}{\partial (t_1 - t_2)} \,
\frac{\partial}{\partial (t_2 - t_1)} \,
\left( -\ii \, D_{ik}(x_1 - x_2) \right) 
\nonumber\\[0.1133ex]
=& \; 
\ii \, \frac{\partial^2}{\partial t^2} \, D_{ik}(x) \,,
\qquad
\qquad
x = x_1 - x_2 \,.
\end{align}
If we work in the mixed representation,
we can implement the differentiation with respect to time
in the Fourier integral as follows,
\begin{align}
\label{sum1}
D^E_{ik}(t, \vec R) =& \;
\langle 0 |
\calT E_i(\vec R_A, t_1) \, E_k(\vec R_B, t_2) 
| 0 \rangle
\nonumber\\[0.1133ex]
=& \; -\ii \, \int \frac{\dd \omega}{2 \pi} \, 
\omega^2 \, D_{ik}(\omega, \vec R) \, \ee^{-\ii \omega t} \,.
\end{align}
Now, let us proceed to the
time-ordered product of dipole operators,
which is given as follows (for atom $A$),
\begin{align}
\alpha_{A,ik}(t_1 - t_2) = & \;
\frac\ii\hbar \, \left< \psi_A \left| 
\rmT( d_{Ai}(t_1) \, d_{Ak}(t_2) ) \right| \psi_A \right> 
\nonumber\\[0.1133ex]
=& \; \frac\ii\hbar \, \left< \psi_A \left| 
\rmT( d_{Ai}(t_1 - t_2) \, d_{Ak}(0) ) \right| \psi_A \right> \,,
\end{align}
and analogously for atom $B$.

Now, according to the prescription that 
Fourier transformation is a summation over 
exponentials with frequency factors
$\exp(-\ii \omega t)$,
\begin{equation}
\alpha_{A,ik}(t) = \int_{-\infty}^\infty 
\frac{\dd \omega}{2 \pi} \, \ee^{-\ii \omega t}  \,
\alpha_{A,ik}(\omega) \,,
\end{equation}
we write
\begin{multline}
\label{sum3}
\left< \psi_A \left| \rmT \; d_{Ai}(t_1) \, d_{Ak}(t_2) \right| \psi_A \right> 
=  -\ii\hbar \, \alpha_{A,ik}(t_1 - t_2) 
\\[0.1133ex]
= -\ii\hbar \, \int_{-\infty}^\infty
\frac{\dd \omega}{2 \pi} \, \ee^{-\ii \omega (t_1 - t_2)}  \,
\alpha_{A,ik}(\omega) \,.
\end{multline}
The time-ordered product of dipole operators can be evaluated 
in terms of the polarizability of the atom,
with the poles being displaced according to the 
Feynman prescription (so that the integrals converge),
\begin{align}
\label{alphaFEYNMAN}
\alpha_{A,ik}(\omega) =& \;
\int_{-\infty}^\infty \dd t \, 
\ee^{\ii \omega t} \, \alpha_{A,ik}(t) 
\nonumber\\[0.1133ex]
=& \;
\frac{\ii}{\hbar}\, \sum_{v} 
\int\limits_{0}^\infty \dd t \,
\ee^{-\frac\ii\hbar (E_v - E_{A} - \hbar \omega -\ii \epsilon) \, t} \,
\nonumber\\[0.1133ex]
& \; \times 
\left< \psi_A \left| d_{Ai} \right| v_A \right> \,
\left< v_A \left| d_{Ak} \right| \psi_A \right>
\nonumber\\[0.1133ex]
& \; + \frac{\ii}{\hbar}\, \sum_{v} 
\int\limits_{-\infty}^0 \dd t \, 
\ee^{\frac\ii\hbar (E_v - E_{A} + \hbar \omega -\ii \epsilon) \, t} \,
\nonumber\\[0.1133ex]
& \; \times \left< \psi_A \left| d_{Ak} \, \right| v_A \right> \,
\left< v_A \left| d_{Ai} \right| \psi_A \right>
\nonumber\\[0.1133ex]
=& \; \sum_{v_A}
\left(
\frac{ \left< \psi_A \left| d_{Ai} \right| v_A \right> \,
\left< v_A \left| d_{Ak} \right| \psi_A \right> }%
{E_{v,A} - \hbar \omega -\ii \epsilon} 
\right.
\nonumber\\[0.1133ex]
& \; \left. + \frac{ \left< \psi_A \left| d_{Ai} \right| v_A \right> \,
\left< v_A \left| d_{Ak} \right| \psi_A \right> }%
{E_{v,A} + \hbar \omega -\ii \epsilon} 
\right) \,,
\end{align}
where $\epsilon>0$ and
\begin{equation}
E_{v,A} = E_{v_A} - E_{A}  
\end{equation}
is the difference between the virtual-state energy
$E_{v_A}$ and the reference-state energy $E_{A}$ of atom~$A$.
In the last step of Eq.~\eqref{alphaFEYNMAN}, we have used the 
fact that the polarizability has to be purely real rather than 
complex for real driving frequency $\omega$,
thus replacing $\left< \psi_A \left| d_{Ak} \right| v_A \right> \,
\left< v_A \left| d_{Ai} \right| \psi_A \right> \to
\left< \psi_A \left| d_{Ai} \right| v_A \right> \,
\left< v_A \left| d_{Ak} \right| \psi_A \right>$ in the second term.
In assigning the time dependence of the 
atomic dipole operators, we have taken into 
account the Heisenberg equation of motion,
$\hbar \frac{\dd}{\dd t} \vec d_A(t) =
\ii \, [H_A, \vec d_A(t)]$, 
where  $H_A$ is the Schr\"{o}dinger Hamiltonian of atom $A$.
The poles in the polarizability $\alpha_{A,ik}$
are displaced according to the Feynman prescription.
Poles occur at 
$\hbar \omega = E_{v,A} - \ii \epsilon$ and at 
$\hbar \omega = -E_{v,A} + \ii \epsilon$.
If the virtual state is displaced toward lower 
energy, i.e., $E_{v,A} < 0$, then the pole at 
$\hbar \omega = -E_{v,A} + \ii \epsilon$ 
migrates into the first quadrant of the complex plane.

The ``correct'' prescription for the 
placement of the poles of the energy denominator 
of the polarizability has recently been controversially 
discussed in the literature~\cite{AnDRSt2003,MiBo2004,MiLoBeBa2008,%
WaEtAl2009,InEtAl2011}.
A different prescription, which puts the poles
into the lower half of the complex plane, has recently 
been used in Ref.~\cite{JePa2015epjd1}.
In this latter study, one considers the relative permittivity $\epsilon_r(\omega)$
of a dilute gas and its relation to the dynamic
dipole polarizability $\alpha(\omega)$ of the gas atoms,
\begin{equation}
\label{one}
\epsilon_r(\omega) = 1 + \frac{N_V}{\epsilon_0} \alpha_R(\omega) \,,
\end{equation}
where $ \alpha_R(\omega)$ denotes the polarizability in 
a pole prescription corresponding to the 
retarded Green function, i.e., with a sign change
($-\ii\epsilon\to+\ii\epsilon$) in the 
second term on the right-hand side of Eq.~\eqref{alphaFEYNMAN}.
Furthermore, $N_V$ is the number density of atoms.
These considerations are valid upon an interpretation
of the dielectric constant in terms of the retarded
Green function $G_R$ which describes the relation of the
dielectric displacement $\vec D(\vec r, t)$ to the
electric field $\vec E(\vec r, t)$,
\begin{equation}
\vec D(\vec r, t) =
\epsilon_0 \, \vec E(\vec r, t) +
\epsilon_0 \int_0^\infty \dd \tau \,
G_R(\tau) \, \vec E(\vec r, t - \tau) \,.
\end{equation}
The Fourier transform is
\begin{equation}
G_R(\omega) = \frac{N_V}{\epsilon_0} \; 
\alpha_R(\omega) \,,
\end{equation}
where $\alpha_R(\omega)$ denotes the ``retarded'' 
polarizability.
The retarded prescription is thus required for 
the dielectric function $\epsilon_r(\omega) 
= 1 + G_R(\omega)$. 
The answer to the question regarding the ``correct''
placement of the poles of the 
polarizability~\cite{AnDRSt2003,MiBo2004,MiLoBeBa2008,%
WaEtAl2009,InEtAl2011} thus is as follows:
Namely, there is no universally ``correct''
displacement for the poles from the real axis. 
Instead, the  correct placement depends on the form of the 
Green function represented by the polarizability, in the 
context of a particular application.
If the retarded Green function is needed, then all poles 
should be displaced into the lower half of the complex
plane, while the Feynman prescription is relevant for
the current calculation, in which the time-ordered 
product of dipole operators is sought.
Neither the retarded nor the Feynman prescription
are universally ``correct''; it depends on the 
context in which the calculation is being performed.

We now reformulate Eq.~\eqref{S4_expr1},
with the help of Eqs.~\eqref{sum2} and~\eqref{sum3},
\begin{multline}
\label{S4_expr2}
\langle \psi, 0 | S^{(4)} | \psi, 0 \rangle =
\frac{1}{2\,\hbar^4}
\int \dd t_1 \int \dd t_2 \int \dd t_3 \int \dd t_4 \,
\\[0.1133ex]
\times 
\left( -\ii \int \frac{\dd \omega_1}{2 \pi} \, \omega_1^2 \,
D_{ij}(\omega_1, \vec R) \, \ee^{-\ii \omega_1 (t_1 - t_2)} \right) 
\\[0.1133ex]
\times 
\left( -\ii \int \frac{\dd \omega_2}{2 \pi} \, \omega_2^2 \,
D_{k\ell}(\omega_2, \vec R) \, \ee^{-\ii \omega_2 (t_3 - t_4)} \right) 
\\[0.1133ex]
\times \left( -\ii \int \frac{\dd \omega_3}{2 \pi} \, 
\hbar\,\alpha_{A,ik}(\omega_3) \, 
\ee^{-\ii \omega_3 (t_1 - t_3)} \right) 
\\[0.1133ex]
\times 
\left( -\ii \int \frac{\dd \omega_4}{2 \pi} \, 
\hbar\,\alpha_{B,j\ell}(\omega_4) \, 
\ee^{-\ii \omega_4 (t_2 - t_4)} \right) \,.
\end{multline}
One now carries out the $\dd t_i$ integrations one after the other,
with the results
$\int \dd t_2 \to 2\pi\,\delta(\omega_1 - \omega_4)$,
then $\int \dd t_3 \to 2\pi\,\delta(\omega_2 - \omega_3)$,
and $\int \dd t_4 \to 2\pi\,\delta(\omega_2 + \omega_4)$.
As a result, the condition
$\omega_1 = \omega_4 = -\omega_2 = -\omega_3$ is implemented
in the final result, yielding
\begin{multline}
\langle \psi, 0 | S^{(4)} | \psi, 0 \rangle =
\frac{1}{2 \hbar^2}
\int \dd t_1 \int \frac{\dd \omega_1}{2 \pi} 
\omega_1^2 (-\omega_1)^2 
\\[0.1133ex]
\times 
D_{ij}(\omega_1, \vec R) \,
D_{k\ell}(-\omega_1, \vec R) \,
\alpha_{A,ik}(-\omega_1) \,
\alpha_{B,j\ell}(\omega_1) 
\\[0.1133ex]
= \frac{T}{2\,\hbar^2}
\int \frac{\dd \omega}{2 \pi} \,
\omega^4 \, D_{ij}(\omega, \vec R) \,
D_{k\ell}(\omega, \vec R) \,
\\[0.1133ex]
\times \alpha_{A,ik}(\omega) \, \alpha_{B,j\ell}(\omega) \,,
\end{multline}
where we use the invariance of the photon 
propagator and of the polarizability 
under the transformation $\omega \leftrightarrow -\omega$ 
[see Eqs.~(\ref{sum2}) and (\ref{alphaFEYNMAN})];
we reemphasize that 
this invariance only holds if the Feynman prescription is used.

%
%
\subsection{Energy Shift}

Using Eq.~\eqref{match}, we obtain the 
diagonal matrix element of the 
effective Hamiltonian,
and thus, the direct term of the 
energy shift $\Delta E^{\dir}$,  as
\begin{subequations}
\label{Ures}
\begin{align}
\Delta E^{\dir} =& \; 
\langle \psi_A \psi_B | H_{\rm eff} | \psi_A \psi_B \rangle
\nonumber\\[2ex]
=& \; \frac{\ii}{2\,\hbar}
\int_{-\infty}^\infty \frac{\dd \omega}{2 \pi} \,
\omega^4 \, D_{ij}(\omega, \vec R)\, D_{k\ell}(\omega, \vec R) \,
\nonumber\\[2ex]
& \; \times \,
\alpha_{A,ik}(\omega) \, \alpha_{B,j\ell}(\omega) \,.
\end{align}
This general result can be applied 
to states of arbitrary symmetry, and is not restricted 
to ground-state atoms.
Invoking the full symmetry of the integrand under a 
sign change of $\omega$, one may write
\begin{align}
\label{DeltaE}
\Delta E^{\dir} =& \; \frac\ii\hbar \int_{0}^\infty \frac{\dd \omega}{2 \pi} \,
\omega^4 \, D_{ij}(\omega, \vec R) \,
D_{k\ell}(\omega, \vec R) \,
\nonumber\\[0.1133ex]
& \; \times \alpha_{A,ik}(\omega) \, \alpha_{B,j\ell}(\omega) \,.
\end{align}
For convenience, we recall the definition of $D_{ij}(\omega, \vec R)$
according to Eq.~\eqref{sum2},
and the definition of $\alpha_{A,ij}(\omega)$ according to
Eq.~\eqref{alphaFEYNMAN},
\begin{align}
\label{CONVENIENCE}
D_{ij}(\omega, \vec R) =& \frac{\hbar \ee^{\ii \frac{|\omega|}{c} R} }%
{4\pi\epsilon_0 c^2 \, R}
\left[ \alpha_{ij} + \beta_{ij}
\left[ \frac{\ii c}{|\omega| R} -
\frac{c^2}{\omega^2 \, R^2} \right] \right] \,,
\nonumber\\[0.1133ex]
\alpha_{ij} =& \; \delta_{ij} - \frac{R_i \, R_j}{R^2} \,,
\qquad
\beta_{ij} = \delta_{ij} - 3 \frac{R_i \, R_j}{R^2} \,,
\nonumber\\[0.1133ex]
\alpha_{A,ij}(\omega)
=& \; \sum_{v_A}
\left(
\frac{ \left< \psi_A \left| d_{Ai} \right| v_A \right> \,
\left< v_A \left| d_{Aj} \right| \psi_A \right> }%
{E_{v,A} - \hbar \omega -\ii \epsilon} 
\right.
\nonumber\\[0.1133ex]
& \; \left. +
\frac{ \left< \psi_A \left| d_{Aj} \right| v_A \right> \,
\left< v_A \left| d_{Ai} \right| \psi_A \right> }%
{E_{v,A} + \hbar \omega -\ii \epsilon}
\right) \,.
\end{align}
Of course, the tensor structures $\alpha_{ij}$ and $\beta_{ij}$ 
need to be distinguished 
from the polarizabilities $\alpha_A$ and $\alpha_B$. 
\end{subequations}

It is a feature of the time-ordered product of
dipole and field operators that all possible time
orderings in time-ordered perturbation theory
(see Fig.~1 of Ref.~\cite{PoTh1995})
are automatically taken into account using a
single propagator.

%

%
%
\subsection{Mixing Term}

In the case of two identical atoms, 
an additional interaction energy term exists 
which needs to be taken into account.
Here, the states $| \psi_A \rangle$ and 
$| \psi_B \rangle$ are obviously not tied to any 
of the atoms, but rather, atom $A$ may assume 
state $| \psi_B \rangle$, and atom $B$ may assume
state $| \psi_A \rangle$ after the interaction.
The eigenstates of the \vdw{} Hamiltonian in
this case are states of the form
$(1/\sqrt{2}) \, \left( 
| \psi_A, \psi_B \rangle \pm
| \psi_B, \psi_A \rangle \right)$
with an energy 
\begin{equation}
\Delta E = \Delta E^\dir \pm \Delta E^\mix \,,
\end{equation}
where $\Delta E^\dir$ is given by Eq.~\eqref{Ures},
and $\Delta E^\mix$ is obtained by calculating
the $S$-matrix element of an initial state 
$| \psi \rangle = | \psi_A \rangle \otimes | \psi_B \rangle$
and the final state
$| \psi' \rangle = | \psi_B \rangle \otimes | \psi_A \rangle$.
In order to calculate the mixing term,
one repeats all steps leading from Eq.~\eqref{SApBpAB}
to Eq.~\eqref{Ures},
for the out state $| \psi' \rangle$ and the 
in state $| \psi \rangle$. The result is
\begin{align}
\label{DeltaEp}
\Delta E^\mix =& \; \frac\ii\hbar \int_{0}^\infty \frac{\dd \omega}{2 \pi} \,
\omega^4 \, D_{ij}(\omega, \vec R) \,
D_{k\ell}(\omega, \vec R) \,
\nonumber\\[0.1133ex]
& \; \times \alpha_{\underline{A}B,ik}(\omega) \, 
\alpha^*_{A\underline{B},j\ell}(\omega) \,.
\end{align}
The definition of $D_{ij}(\omega, \vec R)$ has been recalled 
in Eq.~\eqref{CONVENIENCE}.
The mixed polarizabilities $\alpha_{\underline{A}B,ij}(\omega)$
and $\alpha_{A\underline{B},ij}(\omega)$ are given as follows,
\begin{align}
\label{Pmixed}
\alpha_{\underline{A}B,ij}(\omega)
=& \; \sum_{v_A}
\left(
\frac{ \left< \psi_A \left| d_{Ai} \right| v_A \right> \,
\left< v_A \left| d_{Aj} \right| \psi_B \right> }%
{E_{v,A} - \hbar \omega -\ii \epsilon} 
\right.
\nonumber\\[0.1133ex]
& \; \left. +
\frac{ \left< \psi_A \left| d_{Aj} \right| v_A \right> \,
\left< v_A \left| d_{Ai} \right| \psi_B \right> }%
{E_{v,A} + \hbar \omega -\ii \epsilon}
\right) \,,
\nonumber\\[0.1133ex]
\alpha_{A\underline{B},ij}(\omega)
=& \; \sum_{v_B}
\left(
\frac{ \left< \psi_A \left| d_{Bi} \right| v_B \right> \,
\left< v_B \left| d_{Bj} \right| \psi_B \right> }%
{E_{v,B} - \hbar \omega -\ii \epsilon}
\right.
\nonumber\\[0.1133ex]
& \; \left. +
\frac{ \left< \psi_A \left| d_{Bj} \right| v_B \right> \,
\left< v_B \left| d_{Bi} \right| \psi_B \right> }%
{E_{v,B} + \hbar \omega -\ii \epsilon}
\right) \,.
\end{align}
Here, the designations of the dipole transition operators
in regard to the atoms $A$ and $B$, i.e., 
as $d_{Ai}$ and $d_{Bi}$, constitute mere 
conveniences; for the mixing term to exist,
the two atoms have to be identical and $| \psi_A \rangle$
and $| \psi_B \rangle$ are different states of the 
same atom.
The important feature which differentiates
$\alpha_{\underline{A}B,ij}(\omega)$ from
$\alpha_{A\underline{B},ij}(\omega)$, in the 
case of identical atoms, is the different 
reference state energy in the denominator.

%
%
\section{Applications}
\label{sec3}

%
%
\subsection{Ground--State Interaction}
\label{sec31}

For a reference $S$ state of atom $A$,
denoted as $|\psi_A\rangle = |n_A S\rangle$,
the polarizability tensor assumes the form
\begin{align}
\label{alphaS}
\alpha_{A,ik}(\omega)
=& \; \frac{\delta_{ik}}{3} \, \sum_{v_A}
\left(
\frac{ \langle n_A S | \vec d_A | v_A P \rangle \cdot
\langle v_AP | \vec d_A | n_A S \rangle }%
{E_{v,A} - \hbar \omega -\ii \epsilon} 
\right.
\nonumber\\[0.1133ex]
& \; \left. + \frac{ \langle n_A S | \vec d_A | v_A P \rangle \cdot
\langle v_A P | \vec d_A | n_A S \rangle }%
{E_{v,A} + \hbar \omega -\ii \epsilon}
\right)\nonumber\\[0.1133ex]
& =  \delta_{ik} \; \alpha_A(\omega) \,,
\end{align}
where we denote $S$ and $P$ states by their 
respective symmetry [in this case, 
$E_{v,A} = E_{v_A} - E_A = E(v_A \, P) - E(n_A S)$],
where the reference state energy is that of the 
$S$ state with principal quantum number $n_A$.
This leads to the following tensor structure 
in Eq.~\eqref{Ures},
\begin{multline}
D_{ij}(\omega, \vec R) \, D_{ij}(\omega, \vec R) =
\left(\frac{\hbar}{4 \pi \epsilon_0c^2}\right)^2
\frac{2 \, \ee^{2 \ii \frac{|\omega|}{c} R}}{R^2} \,
\nonumber\\[0.1133ex]
\times \left( 1 + \frac{2 \ii\, c}{|\omega| \, R}
- \frac{5c^2}{(\omega R)^2}
- \frac{6 \ii\,c^3}{(|\omega| R)^3}
+ \frac{3c^4}{(\omega R)^4} \right) \,.
\end{multline}
A Wick rotation of expression~\eqref{Ures} 
then leads to
\begin{align}
\Delta E^\dir =&\;  \; -\frac{\hbar}{\pi c^4 (4 \pi \epsilon_0)^2}
\int\limits_0^\infty
\dd \omega \, \ee^{-2 \omega R/c} \, 
\frac{\omega^4}{R^2} \, 
\nonumber\\[0.1133ex]
& \; \times \left( 1 + \frac{2c}{\omega \, R}
+ \frac{5c^2}{(\omega R)^2}
+ \frac{6c^3}{(\omega R)^3}
+ \frac{3c^4}{(\omega R)^4} \right) 
\nonumber\\[0.1133ex]
& \; \times \alpha_{A}(1S; \ii \omega) \,
\alpha_{B}(1S; \ii \omega) \,,
\end{align}
where we indicate the 
atomic states relevant to the investigation,
for clarity. Expression~\eqref{DeltaE}
verifies known results 
(see Chap.~85 of Ref.~\cite{BeLiPi1982vol4}).

%
%
\subsection{Van der Waals (Close--Range) Limit}
\label{sec32}

A classic result which needs to be 
verified is the close-range limit.
For $R \ll c/\omega$, where $\omega$ is a 
typical transition wavelength, we find
from the dominant term in Eq.~\eqref{sum2}
in this limit,
\begin{equation}
\label{DijCLOSE}
D_{ij}(\omega, \vec R) \approx
- \frac{\hbar}{4 \pi \epsilon_0}\frac{\beta_{ij}}{\omega^2 \, R^3}  \,.
\end{equation}
For arbitrary angular symmetry of the reference state, 
we thus have
\begin{equation}
\Delta E^\dir \approx 
\frac{\ii \hbar \, \beta_{ij} \, \beta_{k\ell}}%
{2 \, (4 \pi \epsilon_0)^2 \, R^6}\,
\int\limits_{-\infty}^{\infty} \frac{\dd \omega}{2 \pi} \,
\alpha_{A,ik}(\omega) \, \alpha_{B,j\ell}(\omega) \,,
\end{equation}
where it is advantageous to keep the integration 
limits as $-\infty$ and $\infty$.
In view of the general result
\begin{multline} \label{eq:IntId}
\hbar\int\limits_{-\infty}^{\infty} \dd \omega 
\left( \sum_{\pm} \frac{1}{E_{v,A} \pm \hbar\omega - \ii \epsilon} \right) 
\left( \sum_{\pm} \frac{1}{E_{v,B} \pm \hbar\omega - \ii \epsilon} \right)
\\
=\frac{4 \pi \ii}{E_{v,A} + E_{v,B}} \,,
\end{multline}
we have
\begin{align}
\label{VDWres}
\Delta E^\dir \approx & \; -
\frac{1}{(4 \pi \epsilon_0)^2}
\frac{\beta_{ij} \, \beta_{k\ell}}{R^6} \,
\sum_{v_A} \sum_{q_B} 
\frac{1}{E_{v,A} + E_{q,B}} 
\nonumber\\[0.1133ex]
& \; \times \left< \psi_A \left| d_{Ai} \right| v_A \right>
\left< v_A \left| d_{Ak} \right| \psi_A  \right>
\nonumber\\[0.1133ex]
& \; \times \left< \psi_B \left| d_{Bj} \right| q_B \right>
\left< q_B \left| d_{B\ell} \right| \psi_B \right> \,.
\end{align}
We denote the virtual states of atom $B$ as $|q \rangle$
as opposed to $|v\rangle$.
This is precisely the expression which would be obtained using second 
order perturbation theory with the \vdw{} potential
\begin{equation}
\calV = \frac{1}{4\pi\,\epsilon_0}\frac{ \beta_{ij} \, d_{Ai} \, d_{Bj} }{R^3} \,,
\end{equation}
which can be obtained by expanding the electrostatic 
potential of the bound electrons and protons in  both atoms
in the limit $| \vec r_A |, | \vec r_B | \ll R$.

\begin{figure} [t]
\begin{center}
\begin{minipage}{0.99\linewidth}
\begin{center}
\includegraphics[width=0.8\linewidth]{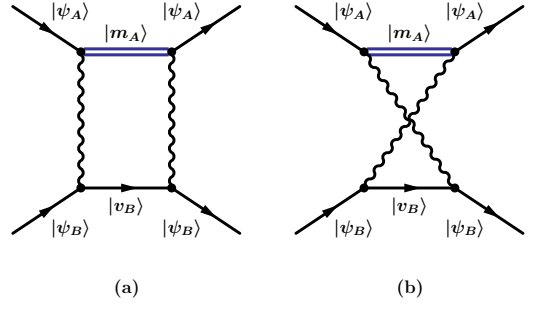}
\caption{\label{fig2}
The virtual resonant contribution due to a lower-lying level $| m_A \rangle$ 
leads to the pole term, which generates the long-range 
interactions for excited states [see Eqs.~\eqref{PresALPHA}
and~\eqref{PmixALPHAp}].}
\end{center}
\end{minipage}
\end{center}
\end{figure}

%
%
\subsection{General Excited Reference States}
\label{sec33}

%
%
\subsubsection{Pole Term}

Let $| m_A \rangle$ be a virtual state of atom $A$,
accessible by a dipole transition,
We now assume that at least one state in atom $A$ is energetically 
lower than the reference state, i.e., $E_{m,A} < 0$,
while atom $B$ is in the ground state.
For the pole term,
in the decomposition~\eqref{alphaFEYNMAN},
we restrict the sum over virtual states $v_A$ 
to just one state whose quantum numbers 
we denote by the multi-index $m_A$ (see Fig.~\ref{fig2}).
A Wick rotation of the integration contour 
$\omega \in (0, \infty)$ from Eq.~\eqref{DeltaE} 
to the imaginary axis
then picks up an additional pole term at 
\begin{equation}
\hbar\omega = -E_{m,A} + \ii \epsilon \,,
\qquad
E_{m,A} < 0 \,,
\end{equation}
which we need to take 
into account. In consequence, the interaction 
energy shift $\Delta E$ due to the energetically
lower virtual state energy with quantum 
numbers $m_A$ (multi-index) naturally splits into a pole 
term $\calQ^\dir_{m_A}$
and a Wick-rotated term $\calW^\dir_{m_A}$,
\begin{equation}
\label{master_separation}
\Delta E^\dir_{m_A} = 
\calQ^\dir_{m_A} + \calW^\dir_{m_A} \,.
\end{equation}
The total direct term is 
\begin{equation}
\label{master_separation1}
\Delta E^\dir =
\left( \sum_{E_{m,A} < 0} \calQ^\dir_{m_A} \right) + \calW^\dir \,,
\end{equation}
where the Wick-rotated term $\calW^\dir$ is obtained after 
the summation over all virtual states (including those
of higher energy) and enters 
the expression in Eq.~\eqref{WICK} below.
For the contribution from the pole,
one finds by Cauchy's residue theorem that 
\begin{subequations}
\begin{widetext}
\begin{align}
\label{PresALPHAexpr} 
\calQ^\dir_{m_A} =& \;
- \mathop{\mbox{Res}}_{\omega = -E_{m,A}/\hbar + \ii \epsilon}
\frac{\omega^4}{\hbar} \, D_{ij}(\omega, \vec R) 
D_{k\ell}(\omega, \vec R) 
\left( \sum_\pm
\frac{ \left< \psi_A \left| d_{Ai} \right| m_A \right> 
\left< m_A \left| d_{Ak} \right| \psi_A \right> }%
{E_{m,A} \pm \hbar \omega -\ii \epsilon} \right)
\alpha_{B,j\ell}(\omega)
\nonumber\\[0.1133ex]
=& \; -\frac{\left< \psi_A \left| d_{Ai} \right| m_A \right> \,
\left< m_A \left| d_{Ak} \right| \psi_A \right>} {(4 \pi \epsilon_0)^2\,R^6} \,
\alpha_{B,j\ell}\left(\frac{E_{m,A}}{\hbar}\right)
\exp\left(-\frac{2 \ii E_{m,A} R}{\hbar c} \right) \,
\left[ 
\beta_{ij} \, \beta_{k\ell} \, \left( 1  
+ 2 \ii \frac{E_{m,A} \, R}{\hbar c} \right)
\right.
\nonumber\\[0.1133ex]
& \; \left. - (2 \alpha_{ij} \, \beta_{k\ell} + \beta_{ij} \, \beta_{k \ell}) 
\left(\frac{E_{m,A} R}{\hbar c}\right)^2 
- 2 \ii \alpha_{ij} \, \beta_{k\ell} \left(\frac{E_{m,A} R}{\hbar c}\right)^3 
+ \alpha_{ij} \, \alpha_{k\ell} \left(\frac{E_{m,A} R}{\hbar c}\right)^4 
\right] 
= \calP^\dir_{m_A} - \frac{\ii}{2} \Gamma^\dir_{m_A} \,.
\end{align}
Here, $\calP^\dir_{m_A}$ is the real part of the interaction 
energy, and $\Gamma^\dir_{m_A}$ is the induced width. 
The identification of the width term $\Gamma^\dir_{m_A}$ follows
the general paradigm that a bound-state energy can be written
as $E = {\rm Re} \, E - \frac{\ii}{2} \Gamma$, where
$\Gamma$ is the width.  One obtains
\begin{multline}
\label{PresALPHA}
\calP^\dir_{m_A} =
-\frac{\left< \psi_A \left| d_{Ai} \right| m_A \right> \,
\left< m_A \left| d_{Ak} \right| \psi_A \right>} {(4 \pi \epsilon_0)^2\,R^6} \,
\alpha_{B,j\ell}\left(\frac{E_{m,A}}{\hbar}\right)
\left\{\cos\left(2 \frac{E_{m,A} R}{\hbar c} \right)
\Biggl[  \beta_{ij} \, \beta_{k\ell}
-\left( 2 \alpha_{ij} \, \beta_{k\ell} 
+ \beta_{ij} \, \beta_{k\ell} \right)
\right.
\\
\left.
\times \left(\frac{E_{m,A} \, R}{\hbar c}\right)^2
+ \alpha_{ij} \, \alpha_{k\ell}
\left(\frac{E_{m,A} \, R}{\hbar c}\right)^4
\Biggr]
+ 2 \frac{E_{m,A} R}{\hbar c}\,
\sin\left( 2 \frac{E_{m,A} R}{\hbar c} \right)
\left[ \beta_{ij} \, \beta_{k\ell} -
\alpha_{ij} \, \beta_{k\ell}  
\left(\frac{E_{m,A} R}{\hbar c}\right)^2
\right]
\right\}\,.
\end{multline}
The width term $\Gamma^\dir_{m_A}$ can be
obtained from $\calP^\dir_{m_A}$ by the substitution
\begin{equation}
\cos\left(\frac{2 E_{m,A} R}{\hbar c}\right) \to 
\sin\left(\frac{2 E_{m,A} R}{\hbar c}\right),
\qquad
\sin\left(\frac{2 E_{m,A} R}{\hbar c}\right) \to 
-\cos\left(\frac{2 E_{m,A} R}{\hbar c}\right) \,,
\end{equation}
and an overall factor two. It reads
\begin{multline}
\label{GAMMAresALPHA}
\Gamma^\dir_{m_A} =
- 2\frac{\left< \psi_A \left| d_{Ai} \right| m_A \right> \,
\left< m_A \left| d_{Ak} \right| \psi_A \right>} {(4 \pi \epsilon_0)^2\,R^6} \,
\alpha_{B,j\ell}\left(\frac{E_{m,A}}{\hbar}\right)
\left\{\sin\left(2 \frac{E_{m,A} R}{\hbar c} \right)
\Biggl[  \beta_{ij} \, \beta_{k\ell}
-\left( 2 \alpha_{ij} \, \beta_{k\ell}
+ \beta_{ij} \, \beta_{k\ell} \right)
\right.
\\
\left.
\times \left(\frac{E_{m,A} \, R}{\hbar c}\right)^2
+ \alpha_{ij} \, \alpha_{k\ell}
\left(\frac{E_{m,A} \, R}{\hbar c}\right)^4
\Biggr]
- 2 \frac{E_{m,A} R}{\hbar c}\,
\cos\left( 2 \frac{E_{m,A} R}{\hbar c} \right)
\left[ \beta_{ij} \, \beta_{k\ell} -
\alpha_{ij} \, \beta_{k\ell}
\left(\frac{E_{m,A} R}{\hbar c}\right)^2
\right]
\right\}\,.
\end{multline}
\end{widetext}
The result~\eqref{PresALPHA} is at variance with the 
corresponding result given in Eq.~(14) of Ref.~\cite{GoMLPo1966},
and with Eq.~(4.1) of Ref.~\cite{PoTh1995}.
It is in better agreement with recently published 
results, such as Eq.~(19) of Ref.~\cite{SaKa2015}
and Eq.~(4) of Ref.~\cite{DoGuLa2015}
(provided we average the latter over the interaction
time $T > 2 R/c$).
We have used a symmetry of the integrand according to 
the replacement 
$\alpha_{ij} \, \beta_{k\ell} + 
\alpha_{k\ell} \, \beta_{ij} + 
\beta_{ij} \, \beta_{k\ell} 
\to 2 \alpha_{ij} \, \beta_{k\ell} + 
\beta_{ij} \, \beta_{k\ell}$
in the cosine term in Eq.~\eqref{PresALPHA},
$\alpha_{ij} \, \beta_{k\ell} + \alpha_{k\ell} \, \beta_{ij}
\to 2 \alpha_{ij} \, \beta_{k\ell}$ in the sine term.
This is valid under the same assumptions as those used
in Eq.~\eqref{alphaFEYNMAN}.

Written in terms of a sum over states for atom $B$,
we have
\begin{multline}
\label{PresSUM}
\alpha_{B,j\ell}\left(\frac{E_{m,A}}{\hbar}\right) =
\sum_{q_B}\left< \psi_B \left| d_{Bj} \right| q_B \right> \,
\left< q_B \left| d_{B\ell} \right| \psi_B \right>
\\[0.1133ex]
\times \left( 
\frac{1}{E_{q,B} - E_{m,A}} +
\frac{1}{E_{q,B} + E_{m,A}} 
\right) \,.
\end{multline}
\end{subequations}
The authors of Ref.~\cite{DoGuLa2015} consider 
a situation with two non-identical atoms, which 
have resonance energies $\hbar\omega_A$ and $\hbar\omega_B$ mutually close.
They define $E_{m,A} = -\hbar\omega_A $ (with manifestly 
positive $\omega_A$) and write $E_{q,B} = \hbar\omega_B$,
assume that $\omega_A \approx \omega_B$,
and define $\Delta_{AB} = \hbar\omega_A - \hbar\omega_B$
with $| \Delta_{AB} | \ll \hbar\omega_A, \hbar\omega_B$.
Furthermore, they restrict the sum over virtual 
states in Eq.~\eqref{PresSUM} to the resonant state,
and they keep only the term $1/(E_{m,A} + E_{q,B})$ in 
Eq.~\eqref{PresSUM}, because under their assumptions
[see Eq.~(4) of Ref.~\cite{DoGuLa2015}],
\begin{equation}
\left| \frac{1}{E_{m,A} + E_{q,B}} \right| = 
\left| -\frac{1}{\Delta_{AB}} \right| \gg 
\left| \frac{1}{E_{q,B} - E_{m,A}} \right| 
\approx 
\frac{1}{2\hbar\omega_B} .
\end{equation}
Our result, given in Eq.~\eqref{PresALPHA},
is much more general as it includes nonresonant terms of atom $B$,
which enter the expression $\alpha_{B,j\ell}\left(E_{m,A}/\hbar\right)$,
and thus not restricted to the special case of distinct
atoms with mutually close resonant frequencies.

%
%
\subsubsection{Wick--Rotated Term}

Let us now consider the Wick-rotated term
from Eq.~\eqref{Ures},
which has the following tensor structure,
\allowdisplaybreaks
\begin{align}
\label{WICK}
\calW^\dir =& \; -\frac{1}{\hbar}\int_{0}^\infty \frac{\dd \omega}{2 \pi} \,
\omega^4 \, D_{ij}(\ii \omega, \vec R) \,
D_{k\ell}(\ii \omega, \vec R) \,
\nonumber\\[0.1133ex]
& \; \times \alpha_{A,ik}(\ii \omega) \, \alpha_{B,j\ell}(\ii \omega) 
\nonumber\\[0.1133ex]
=& \; -\frac{\hbar}{(4 \pi \epsilon_0)^2 \, c^4} \,
\int\limits_{0}^\infty \frac{\dd \omega}{2 \pi} 
\ee^{-2 \omega R/c} \, \frac{\omega^4}{R^2} 
\nonumber\\[0.1133ex]
& \; \times \left[ \alpha_{ij} + 
\left( \frac{c}{\omega R} + \frac{c^2}{(\omega R)^2} \right) 
\beta_{ij} \right]
\alpha_{A,ik}(\ii \omega) 
\nonumber\\[0.1133ex]
& \; \times
\left[ \alpha_{k\ell} + 
\left( \frac{c}{\omega  R} + \frac{c^2}{(\omega R)^2} \right) 
\beta_{k\ell} \right]
\alpha_{B,j\ell}(\ii \omega).
\end{align}
Here, the full polarizabilities are to be used;
i.e., the sum over virtual states 
is not restricted to states with 
a lower energy than that of the reference state, for atom $A$.
According to the nonstandard definition~\eqref{nonstandard},
one has 
\begin{equation}
| \ii \omega | = \sqrt{ (\ii \omega)^2 + \ii \epsilon} = 
\ii \omega \,, \qquad \omega > 0 \,,
\end{equation}
and the Wick rotation can be carried out as usual.

It is now crucial to verify that, in the sum of the 
pole term and the Wick-rotated term, the contribution of the 
virtual state $m_A$--which has lower energy than 
$\psi_A$--to the nonretarded \vdw{} energy~\eqref{VDWres}
gives the expected result. 
The Wick rotation performed in Eq.~\eqref{WICK} is not 
``innocent''; within the Wick-rotated integral,
it changes the sign of the contribution 
of the energetically lower state to the \vdw{} energy.
A compensating term is offered by the pole term,
in a way discussed in the following.

First, we approximate Eq.~\eqref{WICK} for close range 
using the asymptotic behavior of
the photon propagator given by Eq.~\eqref{DijCLOSE}.
In view of the general result
\begin{multline} \label{eq:IntIdWithI}
\hbar\int\limits_{-\infty}^{\infty} \dd \omega 
\left( \sum_{\pm} \frac{1}{E_{m,A} \pm \ii\hbar\omega} \right) \;
\left( \sum_{\pm} \frac{1}{E_{q,B} \pm \ii\hbar\omega} \right) 
\\
=\frac{4 \pi\,\mathrm{sgn}\left(E_{m,A}\right) \,
\mathrm{sgn}\left(E_{q,B}\right)}{\left|E_{m,A}\right| + \left|E_{q,B}\right|} \,,
\end{multline}
an evaluation of the Wick-rotated integral in the 
short-range limit leads to 
\begin{align}
\label{Wapprox}
\calW^\dir_{m_A} \mathop{=}^{R \to 0} & \;
\frac{1}{(4 \pi \epsilon_0)^2 R^6} \sum_{q_B} 
\frac{\beta_{ij} \, \beta_{k\ell}}
{| E_{m,A} | + E_{q,B}} \,
\left< \psi_A \left| d_{Ai} \right| m_A \right> \,
\nonumber\\[0.1133ex]
& \; \times 
\left< m_A \left| d_{Ak} \right| \psi_A \right> \,
\left< \psi_B \left| d_{Bj} \right| q_B \right> \,
\left< q_B \left| d_{B\ell} \right| \psi_B \right>
\nonumber\\[0.1133ex]
= & \; \frac{1}{(4 \pi \epsilon_0)^2 R^6} 
\sum_{q_B} \frac{\beta_{ij} \, \beta_{k\ell}}
{-E_{m,A} + E_{q,B}} \,
\left< \psi_A \left| d_{Ai} \right| m_A \right> \,
\nonumber\\[0.1133ex]
& \; \times
\left< m_A \left| d_{Ak} \right| \psi_A \right> 
\left< \psi_B \left| d_{Bj} \right| q_B \right> 
\left< q_B \left| d_{B\ell} \right| \psi_B \right> .
\end{align}
We have assumed that $E_{m,A} < 0$;
the result is {\em not} equal to the 
contribution of the virtual state $m_A$ to the \vdw{} energy~\eqref{VDWres}.
The compensating term is obtained by 
considering the short-range limit of the pole term,
which is obtained from Eq.~\eqref{PresALPHA} in the 
limit $R \to 0$,
\begin{multline}
\label{Papprox}
\calP^\dir_{m_A} \mathop{=}^{R \to 0} 
-\frac{1}{(4 \pi \epsilon_0)^2 \, R^6}
\left( \sum_{q_B, \pm} 
\frac{\beta_{ij} \beta_{k\ell}}{\pm E_{m,A} + E_{q,B}} \right)
\\[0.1133ex]
\times 
\langle \psi_A | d_{Ai} | m_A \rangle 
\langle m_A | d_{Ak} | \psi_A \rangle 
\langle \psi_B | d_{Bj} | q_B \rangle
\langle q_B | d_{B\ell} | \psi_B \rangle.
\end{multline}
For completeness, we also note the short-range asymptotics
of the width term,
\begin{multline}
\label{GAMMAresapprox}
\Gamma^\dir_{m_A} \mathop{=}^{R \to 0}
\frac{2 [E_{m,A}/(\hbar c)]^3}{3 \, (4 \pi \epsilon_0)^2 \, R^3}
\left( \sum_{q_B, \pm}
\frac{\beta_{ij} \beta_{k\ell} - 3 \alpha_{ij} \, \beta_{k\ell}}{\pm E_{m,A} + E_{q,B}} \right)
\\[0.1133ex]
\times
\langle \psi_A | d_{Ai} | m_A \rangle
\langle m_A | d_{Ak} | \psi_A \rangle
\langle \psi_B | d_{Bj} | q_B \rangle
\langle q_B | d_{B\ell} | \psi_B \rangle.
\end{multline}

The sum of the terms in Eqs.~\eqref{Wapprox} 
and~\eqref{Papprox} restores the 
van der Waals limit,
\begin{multline}
\label{eq:BeforeTask}
\Delta E^\dir_{m_A} = 
\calP^\dir_{m_A} + \calW^\dir_{m_A} \mathop{=}^{R \to 0} 
- \frac{1}{(4 \pi \epsilon_0)^2 \, R^6} \\
\times \sum_{q_B}
\frac{\beta_{ij} \, \beta_{k\ell}}
{E_{m,A} + E_{q,B}}
 \langle \psi_A | d_{Ai} | m_A \rangle
\langle m_A | d_{Ak} | \psi_A \rangle 
\\[0.1133ex]
\; \times 
\langle \psi_B | d_{Bj} | q_B \rangle
\langle q_B | d_{B\ell} | \psi_B \rangle.
\end{multline}
This result precisely corresponds to 
what would be expected from
second-order perturbation theory 
if the Hilbert space of atom $A$ were restricted
in the two states $\psi_A$ and $m_A$.
Supplementing the energetically higher 
states $| v_A \rangle$ for atom $A$,
given in the Wick-rotated form 
Eq.~\eqref{WICK}, one restores the full \vdw{} limit.

Let us now turn our attention to the 
long-range limit.
For the $1S$--$1S$ interaction,
the classic result for very large interatomic separation~\cite{CaPo1948}
calls for a Casimir-Polder $1/R^7$ asymptotics.
This is only valid, as we
now argue, if both atoms are in their ground state. Indeed, in this situation,
only the Wick-rotated contribution subsists, and its asymptotics is indeed of
the Casimir-Polder form. In the general case, however, 
for arbitrary tensor structure, we both have the
Wick-rotated term
\begin{multline}
\label{LONGWICK}
\calW^\dir  \; \mathop{=}^{R \to \infty} \;
-\frac{\hbar c}{8\pi}
\frac{\alpha_{A,ik}(0) \alpha_{B,j\ell}(0)}{ (4 \pi \epsilon_0)^2 \, R^7} \,
\\[0.1133ex]
\times \left( 3 \alpha_{ij}\alpha_{kl} 
+ 5 \alpha_{ij}\beta_{kl}
+ 5 \beta_{ij}\beta_{kl} \right)
\end{multline}
and the pole term which has the long-range asymptotics
\begin{multline}
\label{LONGPOLE}
\calP^\dir_{m_A} \; \mathop{=}^{R \to \infty}  \;
-\frac{1}{(4 \pi \epsilon_0)^2 \, R^2} 
\left(\frac{E_{m,A}}{\hbar c} \right)^4 \,
\cos\left(2\frac{E_{m,A} R}{\hbar c}\right) 
\\[0.1133ex]
\times 
\alpha_{ij}\alpha_{kl}\,
\left< \psi_A \left| d_{Ai} \right| m_A \right> 
\left< m_A \left| d_{Ak} \right| \psi_A \right> 
\\[0.1133ex]
\times
\alpha_{B,j\ell}\left(\frac{E_{m,A}}{\hbar}\right) \,.
\end{multline}
The long-range form of the width term reads as 
\begin{multline}
\label{LONGAMMA}
\Gamma^\dir_{m_A} \; \mathop{=}^{R \to \infty}  \;
-\frac{2}{(4 \pi \epsilon_0)^2 \, R^2}
\left(\frac{E_{m,A}}{\hbar c} \right)^4 \,
\sin\left(2\frac{E_{m,A} R}{\hbar c}\right)
\\[0.1133ex]
\times
\alpha_{ij}\alpha_{kl}\,
\left< \psi_A \left| d_{Ai} \right| m_A \right>
\left< m_A \left| d_{Ak} \right| \psi_A \right>
\\[0.1133ex]
\times \alpha_{B,j\ell}\left(\frac{E_{m,A}}{\hbar}\right) \,.
\end{multline}
This result confirms the existence of an extremely long-range 
\vdw{} interaction for excited states.

%
%
\subsubsection{Mixing Terms}

We now need to start from Eq.~\eqref{DeltaEp}
for the mixing term and analyze the pole term
generated for a virtual state of lower energy, in atom $A$,
and the Wick-rotated term, as well as its nonretarded limit.
The mixing term is relevant only 
for identical atoms.
We recall that for identical atoms,
the eigenstates of the \vdw{} Hamiltonian 
are states of the form
$(1/\sqrt{2}) \, \left(
| \psi_A, \psi_B \rangle \pm
| \psi_B, \psi_A \rangle \right)$,
with an energy $\Delta E^\dir \pm \Delta E^\mix$,
where $\Delta E^\dir$ is given by Eq.~\eqref{Ures},
and $\Delta E^\mix$ by Eq.~\eqref{DeltaEp}.
We write the contribution $\Delta E^\mix_{m_A}$ from an 
energetically lower state $|v_A\rangle = | m_A \rangle$ 
with $E_{m,A} < 0$ as
\begin{equation}
\label{master_separationp}
\Delta E^\mix_{m_A} =
\calQ^\mix_{m_A} + \calW^\mix_{m_A} \,.
\end{equation}
The total mixing term is obtained as the sum 
\begin{equation}
\label{master_separation2}
\Delta E^\mix =
\left( \sum_{E_{m,A} < 0} \calQ^\mix_{m_A} \right) + \calW^\mix \,,
\end{equation}
where $\calW^\mix$ is the total mixing term,
summed over all states, energetically lower as well as 
higher.

The generalization of Eq.~\eqref{PresALPHAexpr} 
to the mixed pole term reads as follows,
\begin{multline}
\calQ^\mix_{m_A} =
-\mathop{\mbox{Res}}_{\omega = -E_{m,A}/\hbar + \ii \epsilon}
\frac{\omega^4}{\hbar} \, D_{ij}(\omega, \vec R)
\\
\times D_{k\ell}(\omega, \vec R) \,
\left(
\frac{ \left< \psi_A \left| d_{Ai} \right| m_A \right>
\left< m_A \left| d_{Ak} \right| \psi_B \right> }%
{E_{m,A} - \hbar \omega -\ii \epsilon}
\right.
\\
\left. + \frac{\left< \psi_A \left| d_{Ai} \right| m_A \right>
\left< m_A \left| d_{Ak} \right| \psi_B \right> }%
{E_{m,A} + \hbar \omega -\ii \epsilon}
\right)
\alpha_{A\underline{B},j\ell}(\omega) \,.
\\
= \calP^\mix_{m_A} - \frac{\ii}{2} \Gamma^\mix_{m_A} \,.
\end{multline}
For the pole term generated at 
$\omega = -E_{m,A} + \ii \epsilon$, we need the 
second term in round brackets,
with the result
\begin{subequations}
\begin{widetext}
\begin{align}
\label{PmixALPHAexpr}
\calQ^\mix_{m_A} =& \;
-\frac{\left< \psi_A \left| d_{Ai} \right| m_A \right> \,
\left< m_A \left| d_{Ak} \right| \psi_B \right>} {(4 \pi \epsilon_0)^2\,R^6} \,
\alpha_{A\underline{B},j\ell}\left(\frac{E_{m,A}}{\hbar}\right)
\exp\left(-\frac{2 \ii E_{m,A} R}{\hbar c} \right) \,
\left[
\beta_{ij} \, \beta_{k\ell} \, \left( 1
+ 2 \ii \frac{E_{m,A} \, R}{\hbar c} \right)
\right.
\nonumber\\[0.1133ex]
& \; \left. - (2 \alpha_{ij} \, \beta_{k\ell} + \beta_{ij} \, \beta_{k \ell})
\left(\frac{E_{m,A} R}{\hbar c}\right)^2
- 2 \ii \alpha_{ij} \, \beta_{k\ell} \left(\frac{E_{m,A} R}{\hbar c}\right)^3
+ \alpha_{ij} \, \alpha_{k\ell} \left(\frac{E_{m,A} R}{\hbar c}\right)^4
\right]
= \calP^\mix - \frac{\ii}{2} \Gamma^\mix \,.
\end{align}
The real part of the pole contribution to the mixing term is
\begin{multline}
\label{PmixALPHAp}
\calP^\mix_{m_A}  =
-\frac{\left< \psi_A \left| d_{Ai} \right| m_A \right> 
\left< m_A \left| d_{Ak} \right| \psi_B \right>} {(4 \pi \epsilon_0)^2\,R^6} 
\alpha_{A\underline{B},j\ell}\left(-\frac{E_{m,A}}{\hbar}\right)
\left\{\cos\left(2 \frac{E_{m,A} R}{\hbar c} \right)
\Biggl[  \beta_{ij} \, \beta_{k\ell}
-\left( 2 \alpha_{ij} \, \beta_{k\ell} 
+ \beta_{ij} \, \beta_{k\ell} \right)
\right.
\\
\left.
\times \left(\frac{E_{m,A} \, R}{\hbar c}\right)^2
+ \alpha_{ij} \, \alpha_{k\ell}
\left(\frac{E_{m,A} \, R}{\hbar c}\right)^4
\Biggr]
+ 2 \frac{E_{m,A} R}{\hbar c}\,
\sin\left(2 \frac{E_{m,A} R}{\hbar c} \right)
\left[ \beta_{ij} \, \beta_{k\ell} -
\alpha_{ij} \, \beta_{k\ell} 
\left(\frac{E_{m,A} R}{\hbar c}\right)^2
\right] \right\}\,.
\end{multline}
The corresponding width term is 
\begin{multline}
\label{GAMMAmixALPHAp}
\Gamma^\mix_{m_A}  =
-2 \frac{\left< \psi_A \left| d_{Ai} \right| m_A \right>
\left< m_A \left| d_{Ak} \right| \psi_B \right>} {(4 \pi \epsilon_0)^2\,R^6}
\alpha_{A\underline{B},j\ell}\left(-\frac{E_{m,A}}{\hbar}\right)
\left\{\sin\left(2 \frac{E_{m,A} R}{\hbar c} \right)
\Biggl[  \beta_{ij} \, \beta_{k\ell}
-\left( 2 \alpha_{ij} \, \beta_{k\ell}
+ \beta_{ij} \, \beta_{k\ell} \right)
\right.
\\
\left.
\times \left(\frac{E_{m,A} \, R}{\hbar c}\right)^2
+ \alpha_{ij} \, \alpha_{k\ell}
\left(\frac{E_{m,A} \, R}{\hbar c}\right)^4
\Biggr]
- 2 \frac{E_{m,A} R}{\hbar c}\,
\cos\left(2 \frac{E_{m,A} R}{\hbar c} \right)
\left[ \beta_{ij} \, \beta_{k\ell} -
\alpha_{ij} \, \beta_{k\ell}
\left(\frac{E_{m,A} R}{\hbar c}\right)^2
\right] \right\}\,.
\end{multline}
\end{widetext}
\end{subequations}

The mixed polarizability $\alpha_{A\underline{B},j\ell}(\omega)$
has been defined in Eq.~\eqref{Pmixed}. The 
(total) Wick-rotated term from Eq.~\eqref{master_separation2} is 
\begin{align}
\label{WICKp}
\calW^\mix =& \; -\frac{1}{\hbar}\int_{0}^\infty \frac{\dd \omega}{2 \pi} \,
\omega^4 \, D_{ij}(\ii \omega, \vec R) \,
D_{k\ell}(\ii \omega, \vec R) \,
\nonumber\\[0.1133ex]
& \; \times \alpha_{A\underline{B},ik}(\ii \omega) \, 
\alpha_{\underline{A}B,j\ell}(\ii \omega) \,.
\end{align}
The generalization of the 
energy shift $\Delta E$ given in Eq.~\eqref{eq:BeforeTask} 
to the mixing term,
in the \vdw{} range, reads as follows,
\begin{multline}
\Delta E^\mix_{m_A} \mathop{=}^{R \to 0} \;
- \frac{1}{(4 \pi \epsilon_0)^2 \, R^6}
\sum_{q_B}
\frac{\beta_{ij} \, \beta_{k\ell}}
{E_{m,A} + E_{q,B}}
\\[0.1133ex]
 \; \times
 \langle \psi_A | d_{Ai} | m_A \rangle
\langle m_A | d_{Bk} | \psi_B \rangle
\langle \psi_A | d_{Aj} | q_B \rangle
\langle q_B | d_{B\ell} | \psi_B \rangle.
\end{multline}
The mixing contribution to the width term,
for close range, is
\begin{multline}
\label{GAMMAmixapprox}
\Gamma^\mix_{m_A} \mathop{=}^{R \to 0}
\frac{2 [E_{m,A}/(\hbar c)]^3}{3 \, (4 \pi \epsilon_0)^2 \, R^3}
\left( \sum_{q_B, \pm}
\frac{\beta_{ij} \beta_{k\ell} - 3 \alpha_{ij} \, \beta_{k\ell}}{\pm E_{m,A} + E_{q,B}} \right)
\\[0.1133ex]
\times
\langle \psi_A | d_{Ai} | m_A \rangle
\langle m_A | d_{Bk} | \psi_B \rangle
\langle \psi_A | d_{Aj} | q_B \rangle
\langle q_B | d_{B\ell} | \psi_B \rangle.
\end{multline}
In the long-range limit, the mixed Wick-rotated term is 
\begin{align}
\label{LONGWICKp}
\calW^\mix \, \mathop{=}^{R \to \infty} & \,
-\frac{\hbar c}{8\pi (4 \pi \epsilon_0)^2 R^7}
\left( 3\alpha_{ij}\alpha_{kl} + 5 \alpha_{ij}\beta_{kl} 
\right.
\nonumber\\[0.1133ex]
& \; \left. 
+ 5\beta_{ij}\beta_{kl} \right) \,
\alpha_{\underline{A}B,ik}(0) \alpha_{A\underline{B},j\ell}(0) \,.
\end{align}
The mixed pole term has the leading long-range asymptotics
\begin{multline}
\label{LONGPOLEp}
\calP^\mix_{m_A} \; \mathop{=}^{R \to \infty} \;
-\frac{1}{(4 \pi \epsilon_0)^2 \, R^2}
\left(\frac{E_{m,A}}{\hbar c} \right)^4 \,
\cos\left(\frac{ 2 E_{m,A} R}{\hbar c} \right) \,
\\[0.1133ex]
\times 
\alpha_{ij}\alpha_{kl}\,
\left< \psi_A \left| d_{Ai} \right| m_A \right>
\left< m_A \left| d_{Ak} \right| \psi_B \right> 
\\[0.1133ex]
\times 
\alpha_{A\underline{B},j\ell}\left(-\frac{E_{m,A}}{\hbar}\right).
\end{multline}
Finally, the mixed width term is
\begin{multline}
\label{LONGGAMMAp}
\Gamma^\mix_{m_A} \; \mathop{=}^{R \to \infty} \;
-\frac{2}{(4 \pi \epsilon_0)^2 \, R^2}
\left(\frac{E_{m,A}}{\hbar c} \right)^4 \,
\sin\left(\frac{ 2 E_{m,A} R}{\hbar c} \right) \,
\\[0.1133ex]
\times
\alpha_{ij}\alpha_{kl}\,
\left< \psi_A \left| d_{Ai} \right| m_A \right>
\left< m_A \left| d_{Ak} \right| \psi_B \right>
\\[0.1133ex]
\times
\alpha_{A\underline{B},j\ell}\left(-\frac{E_{m,A}}{\hbar}\right).
\end{multline}
Due to the symmetry of the wave function,
the total interaction energy $\Delta E^\dir \pm \Delta E^\mix$,
as well as the Wick-rotated term 
\begin{equation}
\calW = \calW^\dir  \pm \calW^\mix
\end{equation}
and the pole and width terms, 
\begin{equation}
\calP_{m_A}  = \calP^\dir_{m_A}  \pm \calP^\mix_{m_A}  \,,
\qquad
\Gamma_{m_A}  = \Gamma^\dir_{m_A}  \pm \Gamma^\mix_{m_A}  \,,
\end{equation}
are the sums of the direct 
and an exchange (mixing) contributions.

%
%
\subsection{Excited Reference $\maybebm{S}$ States}
\label{sec34}

%
%
\subsubsection{Pole Term for $\maybebm{S}$ States}

For $S$ states (i.e., when atom $A$
is in a state with $S$ symmetry), a number of simplifications are possible,
as we can replace $\alpha_{A,ik}(\omega) \to
\delta_{ik} \, \alpha_A(\omega)$ 
[see Eq.~\eqref{alphaS}].
We restrict the discussion to the direct term.
The interaction energy~\eqref{Ures} becomes
\begin{align}
\Delta E^\dir =& \; \frac\ii\hbar \int\limits_0^\infty \frac{\dd \omega}{2 \pi} \,
\omega^4 \, 
D_{ij}(\omega, \vec R) 
D_{ji}(\omega, \vec R) 
\alpha_A(\omega) \, \alpha_B(\omega).
\end{align}
The pole term for an energetically lower $|m_A P\rangle$ state 
becomes
\begin{multline}
\calQ^\dir =
-\frac{2}{3 (4 \pi \epsilon_0)^2 R^6}
\langle n_AS | \vec d_{A} | m_AP \rangle \cdot
\langle m_AP | \vec d_A | n_AS  \rangle
\\[0.1133ex]
\times
\alpha_B\left(\frac{E_{m,A}}{\hbar}\right) \,
\exp\left(-\frac{2 \ii E_{m,A} R}{\hbar c}\right)
\\[0.1133ex]
\times
\left[ 3 + 6 \ii \frac{E_{m,A} R}{\hbar c} 
- 5 \left( \frac{E_{m,A} R}{\hbar c} \right)^2 \right.
\\[0.1133ex]
\left. - 2 \ii \left( \frac{E_{m,A} R}{\hbar c} \right)^3
+ \left( \frac{E_{m,A} R}{\hbar c} \right)^4 \right] \,.
\end{multline}
The real part is 
\begin{multline}
\calP^\dir =
-\frac{2}{3 (4 \pi \epsilon_0)^2 R^6}
\langle n_AS | \vec d_{A} | m_AP \rangle \cdot
\langle m_AP | \vec d_A | n_AS  \rangle
\\[0.1133ex]
\times
\alpha_B\left(\frac{E_{m,A}}{\hbar}\right)
\left\{\cos\left(\frac{2 E_{m,A} R}{\hbar c}\right) 
\right.
\\[0.1133ex]
\times 
\left( 3 - 5 \left(\frac{E_{m,A} R}{\hbar c}\right)^2 + 
\left(\frac{E_{m,A} R}{\hbar c}\right)^4 \right)
\\[0.1133ex]
\left.+ \frac{2 E_{m,A} R}{\hbar c}
\sin\left( \frac{2 E_{m,A} R}{\hbar c} \right) 
\left( 3 - \left(\frac{E_{m,A} R}{\hbar c} \right)^2 \right) \right\}\,.
\end{multline}
The corresponding width term is
\begin{multline}
\Gamma^\dir =
-\frac{4}{3 (4 \pi \epsilon_0)^2 R^6}
\langle n_AS | \vec d_{A} | m_AP \rangle \cdot
\langle m_AP | \vec d_A | n_AS  \rangle
\\[0.1133ex]
\times
\alpha_B\left(\frac{E_{m,A}}{\hbar}\right)
\left\{\sin\left(\frac{2 E_{m,A} R}{\hbar c}\right)
\right.
\\[0.1133ex]
\times
\left( 3 - 5 \left(\frac{E_{m,A} R}{\hbar c}\right)^2 +
\left(\frac{E_{m,A} R}{\hbar c}\right)^4 \right)
\\[0.1133ex]
\left. - \frac{2 E_{m,A} R}{\hbar c}
\cos\left( \frac{2 E_{m,A} R}{\hbar c} \right)
\left( 3 - \left(\frac{E_{m,A} R}{\hbar c} \right)^2 \right) \right\}\,.
\end{multline}
We recognize a number of prefactors also present 
in Eq.~(19) of Ref.~\cite{SaKa2015}
and recall the definition of the $S$-state polarizability from Eq.~\eqref{alphaS}.
In the sum-over-states representation, the polarizability 
relevant to the pole term reads
\begin{align}
\alpha_B\left(\frac{E_{m,A}}{\hbar}\right)
=& \; \frac13 \sum_{q_B}
\langle n_BS | \vec d_{B} | q_BP \rangle \cdot
\langle q_BP | \vec d_B | n_BS  \rangle 
\nonumber\\[0.1133ex]
& \; \times \left( \frac{1}{E_{q,B}-E_{m,A}} +
\frac{1}{E_{q,B} + E_{m,A}} \right) \,.
\end{align}
We recall that the pole term persists only for $E_{m,A} < 0$.

%
%
\subsubsection{Wick--Rotated Term for $\maybebm{S}$ States}

For $S$ states,
the Wick-rotated term~\eqref{WICK} becomes
\begin{align}
\calW^\dir =& \; -\frac{\hbar}{\pi c^4 (4 \pi \epsilon_0)^2}
\int\limits_0^\infty
\frac{\dd \omega}{\pi} \, \ee^{-2 \omega R/c} \, 
\frac{\omega^4}{R^2} \,
\nonumber\\[0.1133ex]
& \; \times \left( 1 + \frac{2}{\omega \, R}
+ \frac{5c^2}{(\omega R)^2}
+ \frac{6c^3}{(\omega R)^3}
+ \frac{3c^4}{(\omega R)^4} \right) \, 
\nonumber\\[0.1133ex]
& \; \times \alpha_A(\ii \omega) \, \alpha_B(\ii \omega) \,.
\end{align}
Irrespective of whether the virtual state
$| m_A \rangle$ is energetically lower or higher than the 
reference state, the long-range limit of $\calW$ 
due to the virtual $P$ state $|m_AP \rangle$ 
is given as follows,
\begin{multline}
\calW^\dir_{m_A} \approx 
\; -\frac{23}{9 \pi}\frac{\hbar c}{(4\pi\epsilon_0)^2}\frac{1}{R^7} \\
\times\frac{\langle \psi_AS | \vec d_{A} | m_AP \rangle \cdot
\langle m_AP | \vec d_A | \psi_AS  \rangle }{E_{m,A}} \\
\times\sum_{q_B}
\frac{\langle \psi_BS | \vec d_B | q_BP \rangle \cdot
\langle q_BP | \vec d_B | \psi_BS \rangle }{E_{q,B}} \,,
\, R \to \infty \,.
\end{multline}
Restoring the sum over $m_A$, one verifies that 
\begin{align}
\calW^\dir \approx & \; -\frac{23}{4 \pi}
\frac{\hbar c}{(4\pi\epsilon_0)^2}
\frac{1}{R^7} \; \alpha_A(0) \, \alpha_B(0) \,,
\qquad R \to \infty \,,
\end{align}
where the static $S$-state polarizabilities are given by
\begin{subequations}
\begin{align}
\alpha_A(0) =& \;
\frac23 \, \sum_{v_A} 
\frac{\langle \psi_AS | \vec d_{A} | v_AP \rangle \cdot
\langle v_AP | \vec d_A | \psi_AS \rangle }{E_{vn,A}} \,,
\\[0.1133ex]
\alpha_B(0) =& \;
\frac23 \, \sum_{q_B} 
\frac{\langle \psi_BS | \vec d_B | q_BP \rangle \cdot
\langle q_BP | \vec d_B | \psi_BS \rangle }{E_{q,B}} \,.
\end{align}
\end{subequations}

%
%
\section{Conclusions}
\label{sec4}

We have investigated the \vdw{} interaction between two atoms in a general
setting, allowing for one of the (conceivably identical) atoms
to be in an excited state. The expressions obtained are
widely applicable. We employed the Feynman prescription propagators for the
electromagnetic field, a prescription which we saw naturally arises out of
time-dependent perturbation theory.
Time-ordered expectation values of the
atomic dipole operators are used.  Our result~\eqref{Ures}
has been kept in fully tensorial form.
Our derivation can be applied to arbitrary angular symmetry of the
atomic reference states involved.
The general result given in Eq.~\eqref{Ures} allows 
us to split the contribution of an 
energetically lower state $| m_A \rangle$ of the excited atom $A$
into a pole and a width term, given in 
Eqs.~\eqref{PresALPHA} and~\eqref{GAMMAresALPHA}, and a
Wick-rotated term, given in Eq.~\eqref{WICK}.
For an energetically lower virtual state $| m_A \rangle$, 
the short-range limit of the 
Wick-rotated term has an interesting sign change 
[see Eq.~\eqref{Wapprox}]
and would lead to a repulsive contribution to the 
\vdw{} interaction. 
However, the pole term compensates this unphysical behavior 
and restores the correct short-range limit
[see Eqs.~\eqref{Papprox} and~\eqref{eq:BeforeTask}].
The additional mixing term incurred for identical
atoms is discussed in Eqs.~\eqref{PmixALPHAp},~\eqref{GAMMAmixALPHAp}
and~\eqref{WICKp}.

The formalism used here involves the matching 
of the scattering amplitude to the effective Hamiltonian.
The use of Feynman propagators allows us to 
drastically reduce the number of diagrams
which need to be considered (Fig.~\ref{fig1})
in comparison to time-ordered perturbation theory~\cite{DoGuLa2015,Do2016},
because all the possible time orderings of the electron-photon
vertices are already contained in the Feynman formalism.
The fully retarded result, and the gerade-ungerade mixing term
including all nonresonant states,
is included in one single, coherent formalism.
Indeed, it was the tremendous simplifications incurred 
by the use of Feynman propagators which allowed 
the simplified evaluation of loop integrals 
in the early days of QED~\cite{Sc1961}.

We confirm that for a system involving an atom
in an excited state, the ``retarded'' $1/R^7$ Casimir-Polder
asymptotics~\cite{CaPo1948} is never fully reached. Indeed, this $1/R^7$
behavior originates in the Wick-rotated version of the integral over photon
frequencies, which gives the interaction energy 
[see Eq.~\eqref{LONGWICK} for the general tensorial 
structure of this Wick-rotated long-range limit]. 
However, if one of the atoms (say, atom~$A$)
is excited, then poles in the complex energy plane are picked up 
upon a Wick rotation of the integration contour. These poles correspond to virtual
states energetically lower than the reference state, and therefore are not
present in the ground state. In the large-interatomic separation limit, these
pole contributions exhibit a surprising $1/R^2$ asymptotics
[see Eq.~\eqref{LONGPOLE}]. When the
interatomic distance becomes longer than the wavelength $\hbar c/|E_{m,A}|$
(where $|E_{m,A}|$ is the transition energy between the reference state and a
lower-energy level accessible through a dipole transition), the pole
contribution becomes larger than the Wick-rotated contribution (the latter
corresponding to the usual Casimir-Polder $1/R^7$ asymptotics), 
with the rule of thumb that 
\begin{equation}
\frac{\mathcal{P}}{\mathcal{W}_{m_A}} \sim 
\alpha^5 \left( \frac{R}{a_0} \right)^5 \,,
\end{equation}
in the Casimir--Polder range.
Let us conclude with a few remarks on the 
interaction of a metastable $2S$ state in hydrogen with 
a ground-state atom~\cite{Ch1972,DeYo1973,TaCh1986}.
The $2P_{1/2}$ states are energetically
lower than the reference $2S$ state but 
displaced only by the Lamb shift $\calL$.
Their contribution is suppressed, even in the oscillatory terms,
due to the $E_{m,A}^4 = \calL^4$ prefactor.
In the Lamb shift range 
$R \sim \hbar c/\calL$ (when $R$ becomes 
commensurate with the Lamb shift wavelength), the static polarizability
of the $2S$ state has the Lamb shift in the denominator,
so that the $1/R^7$--Wick-rotated term of the 
interaction energy shift
is of order $1/(\hbar c/\calL)^7 \, (\calL/\hbar c)^{-1} = [\calL/(\hbar c)]^6$.
For $R \sim \hbar c/\calL$, it competes with the oscillatory
term which is of the same order of
magnitude, namely, 
$[\calL/(\hbar c)]^4 / [R/(\hbar c)]^2 = 
[\calL/(\hbar c)]^4 / [\calL/(\hbar c)]^2 = [\calL/(\hbar c)]^6$.
In the given distance range, the interaction energy 
is of order $\alpha^{24} m_e c^2$, where
$m_e$ is the electron mass,
and thus is negligible.
The oscillatory term exists for the $2S$--$1S$ interaction,
but it dominates only for such long distances that 
no drastic surprises can be expected for 
frequency shifts due to long-range interactions,
within high-precision spectroscopy~\cite{MaEtAl2013prl}.
The suppression mainly is due to the smallness of the 
Lamb shift; analogous observations have 
recently been made in Ref.~\cite{Je2015rapid},
where the $2P$ admixtures to a reference $2S$ state
in hydrogen have been calculated for atom-wall 
interactions: A parametrically interesting long-range 
tail has been identified, but it was found to be suppressed 
due to the smallness of the Lamb shift.

The situation is different for highly excited states,
where the energy shift naturally splits 
into a pole term, a width term and a Wick-rotated term.
This is applicable both to the ``direct'' as well 
as the ``mixing'' term [see Eqs.~\eqref{master_separation}
and~\eqref{master_separationp}]. 
Our general results~\eqref{Ures},~\eqref{PresALPHA},~\eqref{GAMMAresALPHA},
and~\eqref{LONGPOLE} are applicable to the ``direct'' term.
The corresponding results, for the mixing term which is 
relevant for \vdw{} interactions among identical 
atoms, can be found in 
Eqs.~\eqref{DeltaEp},~\eqref{PmixALPHAp},~\eqref{GAMMAmixALPHAp},
and~\eqref{LONGPOLEp}.

%
%
\section*{Acknowledgments}

This research was supported by the National Science Foundation
(Grant PHY--1403973).

\appendix

%
%
\section{Significance of Nonresonant States}

We should clarify the relation of our work to other recent
studies~\cite{DoGuLa2015,Be2015} which are 
based on a restricted subset of atomic states, 
for the two atoms participating in the interaction,
and the reference work~\cite{PoTh1995} which uses 
time-ordered perturbation theory.
Let us start with the latter endeavor.
The Feynman propagators [see Eq.~\eqref{alphaFEYNMAN}], 
which are used in our derivation, capture
different time orderings of the electron-photon interactions in one full sweep.
As the propagator captures 
different time orderings of electron-photon interactions
in one single expression,
it was possible in the early days of QED~\cite{Sc1958}
to carry out the so-called virtual 
loop integrals of the vacuum polarization and 
self energy~\cite{Mo1974a,Mo1974b}. Using the Feynman
formalism, the twelve time-ordered
diagrams for the \vdw{} interaction  (given in a number of places in the
literature, including Fig. 1 of~Ref.~\cite{PoTh1995}),
can be replaced by just two diagrams, given in Fig.~\ref{fig1},
which involve Feynman propagators. The latter approach also eliminates
any guesswork on where to place the infinitesimal imaginary parts in the
denominators which determine the location of the poles.

Our result interpolates between the close-range non-retarded
\vdw{} regime, and the long-range tails.  When one
adds the pole term and the Wick-rotated term, in our approach, then one gets
the van der Waals result back, in the close-range limit
[see Eq.~\eqref{eq:BeforeTask}].  In order for this
to happen, one has to include the nonresonant virtual  states into the
formalism right from the start. In the long range, the 
pole term dominates [see Eq.~\eqref{LONGWICK}].
In the \vdw{} limit, on the other hand, all the nonresonant,
virtual states of the atom become relevant.

The alternative approach, as outlined in Refs.~\cite{DoGuLa2015,Be2015},
restricts the discussion to few ``active'' states, namely, to the ground state,
and a single excited states, for each of the atoms.  Based on this
approximation, the quantum dynamics can be formulate within the few-states
approximation~(for an outline of the formalism used, see also
Ref.~\cite{BeMi2004}).  The validity of this treatment is restricted to
non-identical atoms with two close resonances.

Our approach is much more general.
It would be quite difficult, if not impossible, to generalize
the treatment outlined in Refs.~\cite{DoGuLa2015,Be2015}  to an
infinite number of virtual states. 
This endeavor would inevitably result in an infinite number of coupled
differential equations.
Our general formulas, on one hand, capture the tensor
structure of the pole terms due to energetically lower virtual states (the
$1/R^2$ long-range tail) and on the other hand, yield the correct van der Waals
close-range result (proportional to $1/R^6$).


\begin{thebibliography}{30}%
\makeatletter
\providecommand \@ifxundefined [1]{%
 \@ifx{#1\undefined}
}%
\providecommand \@ifnum [1]{%
 \ifnum #1\expandafter \@firstoftwo
 \else \expandafter \@secondoftwo
 \fi
}%
\providecommand \@ifx [1]{%
 \ifx #1\expandafter \@firstoftwo
 \else \expandafter \@secondoftwo
 \fi
}%
\providecommand \natexlab [1]{#1}%
\providecommand \enquote  [1]{``#1''}%
\providecommand \bibnamefont  [1]{#1}%
\providecommand \bibfnamefont [1]{#1}%
\providecommand \citenamefont [1]{#1}%
\providecommand \href@noop [0]{\@secondoftwo}%
\providecommand \href [0]{\begingroup \@sanitize@url \@href}%
\providecommand \@href[1]{\@@startlink{#1}\@@href}%
\providecommand \@@href[1]{\endgroup#1\@@endlink}%
\providecommand \@sanitize@url [0]{\catcode `\\12\catcode `\$12\catcode
  `\&12\catcode `\#12\catcode `\^12\catcode `\_12\catcode `\%12\relax}%
\providecommand \@@startlink[1]{}%
\providecommand \@@endlink[0]{}%
\providecommand \url  [0]{\begingroup\@sanitize@url \@url }%
\providecommand \@url [1]{\endgroup\@href {#1}{\urlprefix }}%
\providecommand \urlprefix  [0]{URL }%
\providecommand \Eprint [0]{\href }%
\providecommand \doibase [0]{http://dx.doi.org/}%
\providecommand \selectlanguage [0]{\@gobble}%
\providecommand \bibinfo  [0]{\@secondoftwo}%
\providecommand \bibfield  [0]{\@secondoftwo}%
\providecommand \translation [1]{[#1]}%
\providecommand \BibitemOpen [0]{}%
\providecommand \bibitemStop [0]{}%
\providecommand \bibitemNoStop [0]{.\EOS\space}%
\providecommand \EOS [0]{\spacefactor3000\relax}%
\providecommand \BibitemShut  [1]{\csname bibitem#1\endcsname}%
\let\auto@bib@innerbib\@empty
\bibitem [{\citenamefont {Donaire}\ \emph {et~al.}(2015)\citenamefont
  {Donaire}, \citenamefont {Gu\'{e}rout},\ and\ \citenamefont
  {Lambrecht}}]{DoGuLa2015}%
  \BibitemOpen
  \bibfield  {author} {\bibinfo {author} {\bibfnamefont {M.}~\bibnamefont
  {Donaire}}, \bibinfo {author} {\bibfnamefont {R.}~\bibnamefont
  {Gu\'{e}rout}}, \ and\ \bibinfo {author} {\bibfnamefont {A.}~\bibnamefont
  {Lambrecht}},\ }\bibfield  {title} {\enquote {\bibinfo {title}
  {\relax{Quasiresonant van der Waals Interaction between Nonidentical
  Atoms}},}\ }\href@noop {} {\bibfield  {journal} {\bibinfo  {journal} {Phys.
  Rev. Lett.}\ }\textbf {\bibinfo {volume} {115}},\ \bibinfo {pages} {033201}
  (\bibinfo {year} {2015})}\BibitemShut {NoStop}%
\bibitem [{\citenamefont {Safari}\ and\ \citenamefont
  {Karimpour}(2015)}]{SaKa2015}%
  \BibitemOpen
  \bibfield  {author} {\bibinfo {author} {\bibfnamefont {H.}~\bibnamefont
  {Safari}}\ and\ \bibinfo {author} {\bibfnamefont {M.~R.}\ \bibnamefont
  {Karimpour}},\ }\bibfield  {title} {\enquote {\bibinfo {title}
  {\relax{Body-Assisted van der Waals Interaction between Excited Atoms}},}\
  }\href@noop {} {\bibfield  {journal} {\bibinfo  {journal} {Phys. Rev. Lett.}\
  }\textbf {\bibinfo {volume} {114}},\ \bibinfo {pages} {013201} (\bibinfo
  {year} {2015})}\BibitemShut {NoStop}%
\bibitem [{\citenamefont {Berman}(2015)}]{Be2015}%
  \BibitemOpen
  \bibfield  {author} {\bibinfo {author} {\bibfnamefont {P.~R.}\ \bibnamefont
  {Berman}},\ }\bibfield  {title} {\enquote {\bibinfo {title}
  {\relax{Interaction energy of nonidentical atoms}},}\ }\href@noop {}
  {\bibfield  {journal} {\bibinfo  {journal} {Phys. Rev. A}\ }\textbf {\bibinfo
  {volume} {91}},\ \bibinfo {pages} {042127} (\bibinfo {year}
  {2015})}\BibitemShut {NoStop}%
\bibitem [{\citenamefont {Milonni}\ and\ \citenamefont
  {Rafsanjani}(2015)}]{MiRa2015}%
  \BibitemOpen
  \bibfield  {author} {\bibinfo {author} {\bibfnamefont {P.~W.}\ \bibnamefont
  {Milonni}}\ and\ \bibinfo {author} {\bibfnamefont {S.~M.~H.}\ \bibnamefont
  {Rafsanjani}},\ }\bibfield  {title} {\enquote {\bibinfo {title}
  {\relax{Distance dependence of two-atom dipole interactions with one atom in
  an excited state}},}\ }\href@noop {} {\bibfield  {journal} {\bibinfo
  {journal} {Phys. Rev. A}\ }\textbf {\bibinfo {volume} {92}},\ \bibinfo
  {pages} {062711} (\bibinfo {year} {2015})}\BibitemShut {NoStop}%
\bibitem [{\citenamefont {Donaire}(2016)}]{Do2016}%
  \BibitemOpen
  \bibfield  {author} {\bibinfo {author} {\bibfnamefont {M.}~\bibnamefont
  {Donaire}},\ }\bibfield  {title} {\enquote {\bibinfo {title} {\relax{Two-atom
  interaction energies with one atom in an excited state: van der Waals
  potentials versus level shifts}},}\ }\href@noop {} {\bibfield  {journal}
  {\bibinfo  {journal} {Phys. Rev. A}\ }\textbf {\bibinfo {volume} {93}},\
  \bibinfo {pages} {052706} (\bibinfo {year} {2016})}\BibitemShut {NoStop}%
\bibitem [{\citenamefont {Jentschura}(2015)}]{Je2015rapid}%
  \BibitemOpen
  \bibfield  {author} {\bibinfo {author} {\bibfnamefont {U.~D.}\ \bibnamefont
  {Jentschura}},\ }\bibfield  {title} {\enquote {\bibinfo {title}
  {\relax{Long-range atom-wall interactions and mixing terms: Metastable
  hydrogen}},}\ }\href@noop {} {\bibfield  {journal} {\bibinfo  {journal}
  {Phys. Rev. A}\ }\textbf {\bibinfo {volume} {91}},\ \bibinfo {pages}
  {010502(R)} (\bibinfo {year} {2015})}\BibitemShut {NoStop}%
\bibitem [{\citenamefont {Gomberoff}\ \emph {et~al.}(1966)\citenamefont
  {Gomberoff}, \citenamefont {McLone},\ and\ \citenamefont
  {Power}}]{GoMLPo1966}%
  \BibitemOpen
  \bibfield  {author} {\bibinfo {author} {\bibfnamefont {L.}~\bibnamefont
  {Gomberoff}}, \bibinfo {author} {\bibfnamefont {R.~R.}\ \bibnamefont
  {McLone}}, \ and\ \bibinfo {author} {\bibfnamefont {E.~A.}\ \bibnamefont
  {Power}},\ }\bibfield  {title} {\enquote {\bibinfo {title}
  {\relax{Long--Range Retarded Potentials between Molecules}},}\ }\href@noop {}
  {\bibfield  {journal} {\bibinfo  {journal} {J. Chem. Phys.}\ }\textbf
  {\bibinfo {volume} {44}},\ \bibinfo {pages} {4148--4153} (\bibinfo {year}
  {1966})}\BibitemShut {NoStop}%
\bibitem [{\citenamefont {Power}\ and\ \citenamefont
  {Thirunamachandran}(1995)}]{PoTh1995}%
  \BibitemOpen
  \bibfield  {author} {\bibinfo {author} {\bibfnamefont {E.~A.}\ \bibnamefont
  {Power}}\ and\ \bibinfo {author} {\bibfnamefont {T.}~\bibnamefont
  {Thirunamachandran}},\ }\bibfield  {title} {\enquote {\bibinfo {title}
  {\relax{Dispersion forces between molecules with one or both molecules
  excited}},}\ }\href@noop {} {\bibfield  {journal} {\bibinfo  {journal} {Phys.
  Rev. A}\ }\textbf {\bibinfo {volume} {51}},\ \bibinfo {pages} {3660--3666}
  (\bibinfo {year} {1995})}\BibitemShut {NoStop}%
\bibitem [{\citenamefont {Safari}\ \emph {et~al.}(2006)\citenamefont {Safari},
  \citenamefont {Buhmann}, \citenamefont {Welsch},\ and\ \citenamefont
  {Dung}}]{SaBuWeDu2006}%
  \BibitemOpen
  \bibfield  {author} {\bibinfo {author} {\bibfnamefont {H.}~\bibnamefont
  {Safari}}, \bibinfo {author} {\bibfnamefont {S.~Y.}\ \bibnamefont {Buhmann}},
  \bibinfo {author} {\bibfnamefont {D.-G.}\ \bibnamefont {Welsch}}, \ and\
  \bibinfo {author} {\bibfnamefont {H.~T.}\ \bibnamefont {Dung}},\ }\bibfield
  {title} {\enquote {\bibinfo {title} {\relax{Body-assisted van der Waals
  interaction between two atoms}},}\ }\href@noop {} {\bibfield  {journal}
  {\bibinfo  {journal} {Phys. Rev. A}\ }\textbf {\bibinfo {volume} {74}},\
  \bibinfo {pages} {042101} (\bibinfo {year} {2006})}\BibitemShut {NoStop}%
\bibitem [{\citenamefont {Berestetskii}\ \emph {et~al.}(1982)\citenamefont
  {Berestetskii}, \citenamefont {Lifshitz},\ and\ \citenamefont
  {Pitaevskii}}]{BeLiPi1982vol4}%
  \BibitemOpen
  \bibfield  {author} {\bibinfo {author} {\bibfnamefont {V.~B.}\ \bibnamefont
  {Berestetskii}}, \bibinfo {author} {\bibfnamefont {E.~M.}\ \bibnamefont
  {Lifshitz}}, \ and\ \bibinfo {author} {\bibfnamefont {L.~P.}\ \bibnamefont
  {Pitaevskii}},\ }\href@noop {} {\emph {\bibinfo {title} {\relax{Quantum
  Electrodynamics, Volume 4 of the Course on Theoretical Physics}}}},\ \bibinfo
  {edition} {2nd}\ ed.\ (\bibinfo  {publisher} {Pergamon Press},\ \bibinfo
  {address} {Oxford, UK},\ \bibinfo {year} {1982})\BibitemShut {NoStop}%
\bibitem [{\citenamefont {Itzykson}\ and\ \citenamefont
  {Zuber}(1980)}]{ItZu1980}%
  \BibitemOpen
  \bibfield  {author} {\bibinfo {author} {\bibfnamefont {C.}~\bibnamefont
  {Itzykson}}\ and\ \bibinfo {author} {\bibfnamefont {J.~B.}\ \bibnamefont
  {Zuber}},\ }\href@noop {} {\emph {\bibinfo {title} {\relax{Quantum Field
  Theory}}}}\ (\bibinfo  {publisher} {McGraw-Hill},\ \bibinfo {address} {New
  York},\ \bibinfo {year} {1980})\BibitemShut {NoStop}%
\bibitem [{\citenamefont {Mohr}\ \emph {et~al.}(1998)\citenamefont {Mohr},
  \citenamefont {Plunien},\ and\ \citenamefont {Soff}}]{MoPlSo1998}%
  \BibitemOpen
  \bibfield  {author} {\bibinfo {author} {\bibfnamefont {P.~J.}\ \bibnamefont
  {Mohr}}, \bibinfo {author} {\bibfnamefont {G.}~\bibnamefont {Plunien}}, \
  and\ \bibinfo {author} {\bibfnamefont {G.}~\bibnamefont {Soff}},\ }\bibfield
  {title} {\enquote {\bibinfo {title} {\relax{QED corrections in heavy
  atoms}},}\ }\href@noop {} {\bibfield  {journal} {\bibinfo  {journal} {Phys.
  Rep.}\ }\textbf {\bibinfo {volume} {293}},\ \bibinfo {pages} {227--372}
  (\bibinfo {year} {1998})}\BibitemShut {NoStop}%
\bibitem [{\citenamefont {Jentschura}\ and\ \citenamefont
  {Keitel}(2004)}]{JeKe2004aop}%
  \BibitemOpen
  \bibfield  {author} {\bibinfo {author} {\bibfnamefont {U.~D.}\ \bibnamefont
  {Jentschura}}\ and\ \bibinfo {author} {\bibfnamefont {C.~H.}\ \bibnamefont
  {Keitel}},\ }\bibfield  {title} {\enquote {\bibinfo {title} {Radiative
  corrections in laser--dressed atoms: Formalism and applications},}\
  }\href@noop {} {\bibfield  {journal} {\bibinfo  {journal} {Ann. Phys.
  (N.Y.)}\ }\textbf {\bibinfo {volume} {310}},\ \bibinfo {pages} {1--55}
  (\bibinfo {year} {2004})}\BibitemShut {NoStop}%
\bibitem [{\citenamefont {Craig}\ and\ \citenamefont
  {Thirunamachandran}(1984)}]{CrTh1984}%
  \BibitemOpen
  \bibfield  {author} {\bibinfo {author} {\bibfnamefont {D.~P.}\ \bibnamefont
  {Craig}}\ and\ \bibinfo {author} {\bibfnamefont {T.}~\bibnamefont
  {Thirunamachandran}},\ }\href@noop {} {\emph {\bibinfo {title}
  {\relax{Molecular Quantum Electrodynamics}}}}\ (\bibinfo  {publisher} {Dover
  Publications},\ \bibinfo {address} {Mineola, NY},\ \bibinfo {year}
  {1984})\BibitemShut {NoStop}%
\bibitem [{\citenamefont {Andrews}\ \emph {et~al.}(2003)\citenamefont
  {Andrews}, \citenamefont {D\'{a}vila~Romero},\ and\ \citenamefont
  {Stedman}}]{AnDRSt2003}%
  \BibitemOpen
  \bibfield  {author} {\bibinfo {author} {\bibfnamefont {D.~L.}\ \bibnamefont
  {Andrews}}, \bibinfo {author} {\bibfnamefont {L.~C.}\ \bibnamefont
  {D\'{a}vila~Romero}}, \ and\ \bibinfo {author} {\bibfnamefont {G.~E.}\
  \bibnamefont {Stedman}},\ }\bibfield  {title} {\enquote {\bibinfo {title}
  {\relax{Polarizability and the resonance scattering of light: Damping sign
  issues}},}\ }\href@noop {} {\bibfield  {journal} {\bibinfo  {journal} {Phys.
  Rev. A}\ }\textbf {\bibinfo {volume} {67}},\ \bibinfo {pages} {055801}
  (\bibinfo {year} {2003})}\BibitemShut {NoStop}%
\bibitem [{\citenamefont {Milonni}\ and\ \citenamefont
  {Boyd}(2004)}]{MiBo2004}%
  \BibitemOpen
  \bibfield  {author} {\bibinfo {author} {\bibfnamefont {P.~W.}\ \bibnamefont
  {Milonni}}\ and\ \bibinfo {author} {\bibfnamefont {Robert~W.}\ \bibnamefont
  {Boyd}},\ }\bibfield  {title} {\enquote {\bibinfo {title} {\relax{Influence
  of radiative damping on the optical-frequency susceptibility}},}\ }\href@noop
  {} {\bibfield  {journal} {\bibinfo  {journal} {Phys. Rev. A}\ }\textbf
  {\bibinfo {volume} {69}},\ \bibinfo {pages} {023814} (\bibinfo {year}
  {2004})}\BibitemShut {NoStop}%
\bibitem [{\citenamefont {Milonni}\ \emph {et~al.}(2008)\citenamefont
  {Milonni}, \citenamefont {Loudon}, \citenamefont {Berman},\ and\
  \citenamefont {Barnett}}]{MiLoBeBa2008}%
  \BibitemOpen
  \bibfield  {author} {\bibinfo {author} {\bibfnamefont {P.~W.}\ \bibnamefont
  {Milonni}}, \bibinfo {author} {\bibfnamefont {R.}~\bibnamefont {Loudon}},
  \bibinfo {author} {\bibfnamefont {P.~R.}\ \bibnamefont {Berman}}, \ and\
  \bibinfo {author} {\bibfnamefont {S.~M.}\ \bibnamefont {Barnett}},\
  }\bibfield  {title} {\enquote {\bibinfo {title} {\relax{Linear
  polarizabilities of two- and three-level atoms}},}\ }\href@noop {} {\bibfield
   {journal} {\bibinfo  {journal} {Phys. Rev. A}\ }\textbf {\bibinfo {volume}
  {77}},\ \bibinfo {pages} {043835} (\bibinfo {year} {2008})}\BibitemShut
  {NoStop}%
\bibitem [{\citenamefont {Wang}\ \emph {et~al.}(2009)\citenamefont {Wang},
  \citenamefont {Li}, \citenamefont {Wang}, \citenamefont {Zhu},\ and\
  \citenamefont {Suhail~Zubairy}}]{WaEtAl2009}%
  \BibitemOpen
  \bibfield  {author} {\bibinfo {author} {\bibfnamefont {Da-Wei}\ \bibnamefont
  {Wang}}, \bibinfo {author} {\bibfnamefont {Ai-Jun}\ \bibnamefont {Li}},
  \bibinfo {author} {\bibfnamefont {Li-Gang}\ \bibnamefont {Wang}}, \bibinfo
  {author} {\bibfnamefont {Shi-Yao}\ \bibnamefont {Zhu}}, \ and\ \bibinfo
  {author} {\bibfnamefont {M.}~\bibnamefont {Suhail~Zubairy}},\ }\bibfield
  {title} {\enquote {\bibinfo {title} {\relax{Effect of the counterrotating
  terms on polarizability in atom-field interactions}},}\ }\href@noop {}
  {\bibfield  {journal} {\bibinfo  {journal} {Phys. Rev. A}\ }\textbf {\bibinfo
  {volume} {80}},\ \bibinfo {pages} {063826} (\bibinfo {year}
  {2009})}\BibitemShut {NoStop}%
\bibitem [{\citenamefont {Intravaia}\ \emph {et~al.}(2011)\citenamefont
  {Intravaia}, \citenamefont {Behunin}, \citenamefont {Milonni}, \citenamefont
  {Ford},\ and\ \citenamefont {O'Connell}}]{InEtAl2011}%
  \BibitemOpen
  \bibfield  {author} {\bibinfo {author} {\bibfnamefont {F.}~\bibnamefont
  {Intravaia}}, \bibinfo {author} {\bibfnamefont {R.}~\bibnamefont {Behunin}},
  \bibinfo {author} {\bibfnamefont {P.~W.}\ \bibnamefont {Milonni}}, \bibinfo
  {author} {\bibfnamefont {G.~W.}\ \bibnamefont {Ford}}, \ and\ \bibinfo
  {author} {\bibfnamefont {R.~F.}\ \bibnamefont {O'Connell}},\ }\bibfield
  {title} {\enquote {\bibinfo {title} {\relax{Consistency of a causal theory of
  radiative reaction with the optical theorem}},}\ }\href@noop {} {\bibfield
  {journal} {\bibinfo  {journal} {Phys. Rev. A}\ }\textbf {\bibinfo {volume}
  {84}},\ \bibinfo {pages} {035801} (\bibinfo {year} {2011})}\BibitemShut
  {NoStop}%
\bibitem [{\citenamefont {Jentschura}\ and\ \citenamefont
  {Pachucki}(2015)}]{JePa2015epjd1}%
  \BibitemOpen
  \bibfield  {author} {\bibinfo {author} {\bibfnamefont {U.~D.}\ \bibnamefont
  {Jentschura}}\ and\ \bibinfo {author} {\bibfnamefont {K.}~\bibnamefont
  {Pachucki}},\ }\bibfield  {title} {\enquote {\bibinfo {title}
  {\relax{Functional form of the imaginary part of the atomic
  polarizability}},}\ }\href@noop {} {\bibfield  {journal} {\bibinfo  {journal}
  {Eur. Phys. J. D}\ }\textbf {\bibinfo {volume} {69}},\ \bibinfo {pages} {118}
  (\bibinfo {year} {2015})}\BibitemShut {NoStop}%
\bibitem [{\citenamefont {Casimir}\ and\ \citenamefont
  {Polder}(1948)}]{CaPo1948}%
  \BibitemOpen
  \bibfield  {author} {\bibinfo {author} {\bibfnamefont {H.~B.~G.}\
  \bibnamefont {Casimir}}\ and\ \bibinfo {author} {\bibfnamefont
  {D.}~\bibnamefont {Polder}},\ }\bibfield  {title} {\enquote {\bibinfo {title}
  {\relax{The Influence of Radiation on the London-van-der-Waals Forces}},}\
  }\href@noop {} {\bibfield  {journal} {\bibinfo  {journal} {Phys. Rev.}\
  }\textbf {\bibinfo {volume} {73}},\ \bibinfo {pages} {360--372} (\bibinfo
  {year} {1948})}\BibitemShut {NoStop}%
\bibitem [{\citenamefont {Schweber}(1961)}]{Sc1961}%
  \BibitemOpen
  \bibfield  {author} {\bibinfo {author} {\bibfnamefont {S.~S.}\ \bibnamefont
  {Schweber}},\ }\href@noop {} {\emph {\bibinfo {title} {\relax{An Introduction
  to Relativistic Quantum Field Theory}}}}\ (\bibinfo  {publisher} {Harper \&
  Row},\ \bibinfo {address} {New York, NY},\ \bibinfo {year}
  {1961})\BibitemShut {NoStop}%
\bibitem [{\citenamefont {Chibisov}(1972)}]{Ch1972}%
  \BibitemOpen
  \bibfield  {author} {\bibinfo {author} {\bibfnamefont {M.~I.}\ \bibnamefont
  {Chibisov}},\ }\bibfield  {title} {\enquote {\bibinfo {title}
  {\relax{Dispersion Interaction of Neutral Atoms}},}\ }\href@noop {}
  {\bibfield  {journal} {\bibinfo  {journal} {Opt. Spectrosc.}\ }\textbf
  {\bibinfo {volume} {32}},\ \bibinfo {pages} {1--3} (\bibinfo {year}
  {1972})}\BibitemShut {NoStop}%
\bibitem [{\citenamefont {Deal}\ and\ \citenamefont {Young}(1973)}]{DeYo1973}%
  \BibitemOpen
  \bibfield  {author} {\bibinfo {author} {\bibfnamefont {W.~J.}\ \bibnamefont
  {Deal}}\ and\ \bibinfo {author} {\bibfnamefont {R.~H.}\ \bibnamefont
  {Young}},\ }\bibfield  {title} {\enquote {\bibinfo {title}
  {\relax{Long--Range Dispersion Interactions Involving Excited Atoms; the
  H(1s)---H(2s) Interaction}},}\ }\href@noop {} {\bibfield  {journal} {\bibinfo
   {journal} {Int. J. Quantum Chem.}\ }\textbf {\bibinfo {volume} {7}},\
  \bibinfo {pages} {877--892} (\bibinfo {year} {1973})}\BibitemShut {NoStop}%
\bibitem [{\citenamefont {Tang}\ and\ \citenamefont {Chan}(1986)}]{TaCh1986}%
  \BibitemOpen
  \bibfield  {author} {\bibinfo {author} {\bibfnamefont {A.~Z.}\ \bibnamefont
  {Tang}}\ and\ \bibinfo {author} {\bibfnamefont {F.~T.}\ \bibnamefont
  {Chan}},\ }\bibfield  {title} {\enquote {\bibinfo {title} {\relax{Dynamic
  Multipole polarizability of atomic hydrogen}},}\ }\href@noop {} {\bibfield
  {journal} {\bibinfo  {journal} {Phys. Rev. A}\ }\textbf {\bibinfo {volume}
  {33}},\ \bibinfo {pages} {3671--3678} (\bibinfo {year} {1986})}\BibitemShut
  {NoStop}%
\bibitem [{\citenamefont {Matveev}\ \emph {et~al.}(2013)\citenamefont
  {Matveev}, \citenamefont {Parthey}, \citenamefont {Predehl}, \citenamefont
  {Alnis}, \citenamefont {Beyer}, \citenamefont {Holzwarth}, \citenamefont
  {Udem}, \citenamefont {Wilken}, \citenamefont {Kolachevsky}, \citenamefont
  {Abgrall}, \citenamefont {Rovera}, \citenamefont {Salomon}, \citenamefont
  {Laurent}, \citenamefont {Grosche}, \citenamefont {Terra}, \citenamefont
  {Legero}, \citenamefont {Schnatz}, \citenamefont {Weyers}, \citenamefont
  {Altschul},\ and\ \citenamefont {H\"{a}nsch}}]{MaEtAl2013prl}%
  \BibitemOpen
  \bibfield  {author} {\bibinfo {author} {\bibfnamefont {A.}~\bibnamefont
  {Matveev}}, \bibinfo {author} {\bibfnamefont {C.~G.}\ \bibnamefont
  {Parthey}}, \bibinfo {author} {\bibfnamefont {K.}~\bibnamefont {Predehl}},
  \bibinfo {author} {\bibfnamefont {J.}~\bibnamefont {Alnis}}, \bibinfo
  {author} {\bibfnamefont {A.}~\bibnamefont {Beyer}}, \bibinfo {author}
  {\bibfnamefont {R.}~\bibnamefont {Holzwarth}}, \bibinfo {author}
  {\bibfnamefont {T.}~\bibnamefont {Udem}}, \bibinfo {author} {\bibfnamefont
  {T.}~\bibnamefont {Wilken}}, \bibinfo {author} {\bibfnamefont
  {N.}~\bibnamefont {Kolachevsky}}, \bibinfo {author} {\bibfnamefont
  {M.}~\bibnamefont {Abgrall}}, \bibinfo {author} {\bibfnamefont
  {D.}~\bibnamefont {Rovera}}, \bibinfo {author} {\bibfnamefont
  {C.}~\bibnamefont {Salomon}}, \bibinfo {author} {\bibfnamefont
  {P.}~\bibnamefont {Laurent}}, \bibinfo {author} {\bibfnamefont
  {G.}~\bibnamefont {Grosche}}, \bibinfo {author} {\bibfnamefont
  {O.}~\bibnamefont {Terra}}, \bibinfo {author} {\bibfnamefont
  {T.}~\bibnamefont {Legero}}, \bibinfo {author} {\bibfnamefont
  {H.}~\bibnamefont {Schnatz}}, \bibinfo {author} {\bibfnamefont
  {S.}~\bibnamefont {Weyers}}, \bibinfo {author} {\bibfnamefont
  {B.}~\bibnamefont {Altschul}}, \ and\ \bibinfo {author} {\bibfnamefont
  {T.~W.}\ \bibnamefont {H\"{a}nsch}},\ }\bibfield  {title} {\enquote {\bibinfo
  {title} {\relax{Precision Measurement of the Hydrogen 1S--2S Frequency via a
  920-km Fiber Link}},}\ }\href@noop {} {\bibfield  {journal} {\bibinfo
  {journal} {Phys. Rev. Lett.}\ }\textbf {\bibinfo {volume} {110}},\ \bibinfo
  {pages} {230801} (\bibinfo {year} {2013})}\BibitemShut {NoStop}%
\bibitem [{\citenamefont {Schwinger}(1958)}]{Sc1958}%
  \BibitemOpen
  \bibfield  {author} {\bibinfo {author} {\bibfnamefont {J.}~\bibnamefont
  {Schwinger}},\ }\href@noop {} {\emph {\bibinfo {title} {\relax{Selected
  Papers on Quantum Electrodynamics}}}}\ (\bibinfo  {publisher} {Dover
  Publications},\ \bibinfo {address} {New York, USA},\ \bibinfo {year}
  {1958})\BibitemShut {NoStop}%
\bibitem [{\citenamefont {Mohr}(1974{\natexlab{a}})}]{Mo1974a}%
  \BibitemOpen
  \bibfield  {author} {\bibinfo {author} {\bibfnamefont {P.~J.}\ \bibnamefont
  {Mohr}},\ }\bibfield  {title} {\enquote {\bibinfo {title}
  {\relax{Self--Energy Radiative Corrections in Hydrogen--Like Systems}},}\
  }\href@noop {} {\bibfield  {journal} {\bibinfo  {journal} {Ann. Phys.
  (N.Y.)}\ }\textbf {\bibinfo {volume} {88}},\ \bibinfo {pages} {26--51}
  (\bibinfo {year} {1974}{\natexlab{a}})}\BibitemShut {NoStop}%
\bibitem [{\citenamefont {Mohr}(1974{\natexlab{b}})}]{Mo1974b}%
  \BibitemOpen
  \bibfield  {author} {\bibinfo {author} {\bibfnamefont {P.~J.}\ \bibnamefont
  {Mohr}},\ }\bibfield  {title} {\enquote {\bibinfo {title} {\relax{Numerical
  Evaluation of the $1S_{1/2}$ Radiative Level Shift}},}\ }\href@noop {}
  {\bibfield  {journal} {\bibinfo  {journal} {Ann. Phys. (N.Y.)}\ }\textbf
  {\bibinfo {volume} {88}},\ \bibinfo {pages} {52--87} (\bibinfo {year}
  {1974}{\natexlab{b}})}\BibitemShut {NoStop}%
\bibitem [{\citenamefont {Berman}\ and\ \citenamefont
  {Milonni}(2004)}]{BeMi2004}%
  \BibitemOpen
  \bibfield  {author} {\bibinfo {author} {\bibfnamefont {P.~R.}\ \bibnamefont
  {Berman}}\ and\ \bibinfo {author} {\bibfnamefont {P.~W.}\ \bibnamefont
  {Milonni}},\ }\bibfield  {title} {\enquote {\bibinfo {title}
  {\relax{Microscopic Theory of Modified Spontaneous Emission in a
  Dielectric}},}\ }\href@noop {} {\bibfield  {journal} {\bibinfo  {journal}
  {Phys. Rev. Lett.}\ }\textbf {\bibinfo {volume} {92}},\ \bibinfo {pages}
  {053601} (\bibinfo {year} {2004})}\BibitemShut {NoStop}%
\end{thebibliography}
\end{document}